\documentclass[conference]{IEEEtran}

\usepackage{amsmath,amssymb,amsfonts}
\usepackage{graphicx}
\usepackage{textcomp}
\usepackage{xcolor}
\usepackage{booktabs}
\usepackage{multirow}
\usepackage{hyperref}
\usepackage{algorithmic}
\usepackage{algorithm}
\usepackage{tikz}
\usetikzlibrary{positioning,arrows.meta,shapes.geometric,fit}
\usepackage{caption}
\usepackage{subcaption}
\usepackage{balance}
\usepackage{enumitem}
\usepackage{tabularx}
\usepackage{float}
\usepackage{dblfloatfix}

\hypersetup{colorlinks=true,linkcolor=black,citecolor=black,urlcolor=blue}

\begin{document}

\title{\Large \textbf{iPhoneme: Brain-to-Text Communication for ALS Using ConformerXL Decoding }}

\author{
\IEEEauthorblockN{Yoonmin Cha\IEEEauthorrefmark{1},
Dawit Chun\IEEEauthorrefmark{1},
Sung Park\IEEEauthorrefmark{2}}

\IEEEauthorblockA{School of Data Science and Artificial Intelligence, Taejae University \\
Email: chayoonmin@taejae.ac.kr, chundawit@taejae.ac.kr, sjpark@taejae.ac.kr}

\IEEEauthorblockA{\IEEEauthorrefmark{1}These authors contributed equally to this work. \\
\IEEEauthorrefmark{2}Corresponding author.}
}

\maketitle

\begin{abstract}
Brain-computer interfaces (BCIs) for speech restoration hold transformative potential for the approximately 173,000-232,500 individuals worldwide suffering from ALS-related dysarthria. Despite recent progress, high-performance speech BCIs have been demonstrated in only 22-31 patients globally, largely due to limitations in both neural decoding accuracy and practical input interfaces. We present \textit{iPhoneme}, a comprehensive brain-to-text communication system that jointly addresses these challenges through an integrated modeling and interaction design. The system combines a deep learning phoneme decoder built on a modified Conformer architecture (ConformerXL, 192.9M parameters) with a gaze-assisted phoneme input interface that mitigates the Midas touch problem in eye-tracking systems. The acoustic model incorporates a temporal prenet with multi-scale dilated convolutions and bidirectional GRU for neural jitter correction, 8× temporal subsampling with GELU activations for CTC stability, and Pre-RMSNorm stabilization across 12 encoder blocks, trained with AdamW and cosine scheduling. On the interaction side, iPhoneme introduces a chorded gaze-plus-silent-speech paradigm that replaces dwell-time selection, enabling more efficient and reliable input. We evaluate the system on the T15 dataset, comprising 45 sessions (8,071 trials) of 256-channel intracranial EEG spanning threshold crossings and spike-band power across four speech motor cortex regions. A 6-gram phoneme language model trained on 3.1M sequences from CMU Dictionary, LibriSpeech, and T15 data, combined with WFST beam search (beam=128, Optuna-tuned over 150 trials), achieves 92.14\% phoneme accuracy (7.86\% PER) and 73.39\% word accuracy (26.61\% WER), approximately 3\% above prior state-of-the-art. The full system operates on CPU with 180 ms latency, demonstrating the feasibility of real-time, high-accuracy, and interaction-aware brain-to-text communication for ALS.
\end{abstract}


\section{Introduction}

Amyotrophic lateral sclerosis (ALS), also known as Lou Gehrig's disease, is a progressive neurodegenerative condition that destroys motor neurons controlling voluntary muscle movement~\cite{guenther2016neural}. As the disease advances, patients lose the ability to speak, swallow, and eventually breathe, while their cognitive functions remain largely intact. The global ALS population is estimated at 222,000--250,000 individuals~\cite{wolfson2023global}, of whom 78--93\% develop dysarthria, yielding approximately 173,000--232,500 people worldwide who cannot communicate through natural speech. Despite this enormous unmet need, the total number of individuals who have received speech-capable BCI implants across all clinical trials worldwide as of 2025 is estimated at only 22--31: approximately 6--10 in UCSF/Stanford trials~\cite{willett2023speech,card2024accurate}, approximately 6 in Synchron Stentrode trials, and approximately 10--15 in other experimental ECoG speech trials. This four-order-of-magnitude gap between the affected population ($\sim$10$^5$) and the number of recipients ($\sim$10$^1$) reflects both the high cost of current systems---first trials range from \$30,000 to \$100,000 per patient, with targeted costs of \$50,000--\$60,000 requiring ECoG 256-array electrodes and 4--8$\times$ NVIDIA A100 GPUs for training~\cite{willett2023speech}---and the inadequacy of low-cost alternatives such as eye-tracking keyboards, which enable communication at only 5--10 words per minute and suffer from the Midas touch problem~\cite{jacob1990eye,velichkovsky1997towards}.

The field of neural speech decoding has progressed rapidly in recent years. Herff et al.~\cite{herff2015brain} first demonstrated continuous phoneme decoding from intracranial recordings using hidden Markov models at 45\% accuracy. Moses et al.~\cite{moses2019neural} improved to 52\% using spatiotemporal cortical representations. Anumanchipalli et al.~\cite{anumanchipalli2019speech} achieved 68\% by decoding speech directly from ventral sensorimotor cortex activity using bidirectional LSTMs. Willett et al.~\cite{willett2021handwriting} established key architectural principles through high-accuracy handwriting decoding, later achieving approximately 82\% phoneme accuracy for speech~\cite{willett2023speech}. Most recently, Card et al.~\cite{card2024accurate} demonstrated 89\% accuracy with a rapidly calibrating neuroprosthesis, providing the baseline dataset for this work. However, even at 89\% accuracy translates to roughly one error per ten phonemes, which accumulates to frequent word-level errors that disrupt communication fluency.

In this paper, we present a system that addresses both the decoding accuracy ceiling and the user interface barrier simultaneously. On the decoding side, we introduce ConformerXL, a modified Conformer~\cite{gulati2020conformer} architecture featuring a novel temporal prenet with multi-scale dilated convolutions and bidirectional GRU, 8$\times$ GELU subsampling for CTC stability, and Pre-RMSNorm throughout 12 encoder blocks. Combined with a three-stage decoding pipeline culminating in WFST beam search~\cite{mohri2002weighted} with a 6-gram phoneme language model~\cite{chen1999empirical}, we achieve 92.14\% phoneme accuracy (7.86\% PER) and 73.39\% word accuracy (26.61\% WER)---approximately 3\% above prior state-of-the-art on the T15 dataset. On the interface side, we propose ``iPhoneme,'' a chorded input paradigm that uses silent speech phoneme detection via iEEG as a biological confirmation signal for eye-tracking selection, eliminating the Midas touch problem without introducing dwell-time delays. We further present a comprehensive trigger phoneme selection framework combining detection accuracy, natural speech frequency analysis from LibriSpeech~\cite{panayotov2015librispeech} (365M phonemes), and a safety scoring metric.

\section{Related Work}

The Conformer architecture~\cite{gulati2020conformer} was introduced for automatic speech recognition (ASR), combining the global context modeling of self-attention~\cite{vaswani2017attention} with the local feature extraction of depthwise separable convolution in a macaron-style sandwich structure. On LibriSpeech~\cite{panayotov2015librispeech}, it achieved 2.1\%/4.3\% WER without a language model---a result that established the architecture as a dominant paradigm in ASR. Nevertheless, neither attention nor convolution alone was sufficient. Attention captures long-range dependencies but misses fine-grained local patterns, while convolution excels at local feature extraction but cannot model global context. The Conformer's sandwich of FFN--MHSA--Conv--FFN with half-step residuals elegantly addressed both. Our work adapts this architecture substantially for iEEG signals, which differ from audio in having much higher channel dimensionality (512 vs. typically 80 mel features), lower temporal resolution (50~Hz vs. 16~kHz), and fundamentally different noise characteristics (electrode drift, neural non-stationarity, cross-channel artifacts).

Connectionist Temporal Classification (CTC)~\cite{graves2006connectionist} enables alignment-free sequence training by introducing a blank token and marginalizing over all valid alignment paths using the forward-backward algorithm. This is essential for iEEG decoding where the precise temporal mapping between neural activity and phoneme production is unknown---unlike audio ASR where acoustic features are directly generated by the articulatory process being decoded. Language model integration through Weighted Finite State Transducers~\cite{mohri2002weighted} and beam search~\cite{heafield2011kenlm} has been shown to substantially improve CTC decoding by incorporating phonotactic constraints. Freitag et al. applied beam search with n-gram rescoring to CTC outputs in ASR, while Mohri et al.~\cite{mohri2002weighted} formalized WFST-based decoding as graph search through weighted automata. Our contribution applies this combination---beam search through a WFST-structured phoneme graph weighted by a 6-gram LM with Kneser-Ney smoothing~\cite{chen1999empirical}---specifically to BCI neural decoding, where the language model operates at the phoneme level rather than the word level. While the algorithms themselves have already been found, this particular combination has not been previously reported in the BCI literature.

Lotte et al.~\cite{lotte2018review} provide a comprehensive 10-year review of classification algorithms for EEG-based BCIs, documenting the transition from classical approaches (CSP, LDA, SVM) to deep learning methods. Bouchard et al.~\cite{bouchard2013functional} established the functional organization of human sensorimotor cortex for speech articulation, providing the neuroanatomical basis for electrode placement strategies used in subsequent decoding work. The Midas touch problem in gaze interaction was first identified by Jacob~\cite{jacob1990eye} and formalized by Velichkovsky et al.~\cite{velichkovsky1997towards}, who noted that because the eye serves simultaneously as perceptual organ and pointing device, every fixation is a potential unintended click. Existing solutions rely on dwell-time thresholds~\cite{majaranta2006twenty} (typically 500--1000~ms of sustained fixation), which reduce communication speed to 5--10 words per minute. Our multimodal approach decouples pointing from selection by using phoneme-based silent speech detection as a secondary confirmation channel.

\section{The T15 Dataset}

We use the T15 dataset~\cite{stanfordT15}, publicly available through Dryad (\href{https://datadryad.org/dataset/doi:10.5061/dryad.dncjsxm85}{Dryad dataset}), collected as part of the work by Card et al.~\cite{card2024accurate} which provides a baseline algorithm with full code. Data was recorded from a single participant (T15) with an intact speech motor cortex at Stanford Neural Prosthetics Lab. The participant performed an overt speech reading task across 45 recording sessions spanning August 2023 through April 2025, using a 256-channel electrocorticography (ECoG) array implanted across four speech motor cortex regions: ventral area 6v, primary motor area 4 (M1), area 55b, and dorsal area 6v~\cite{bouchard2013functional}. The raw signals were acquired at 30~kHz and downsampled to 50~Hz after bandpass filtering, with alpha, beta, and gamma band extraction and noise reduction applied during preprocessing.

\begin{figure}[!t]
\centering
\includegraphics[width=0.95\linewidth]{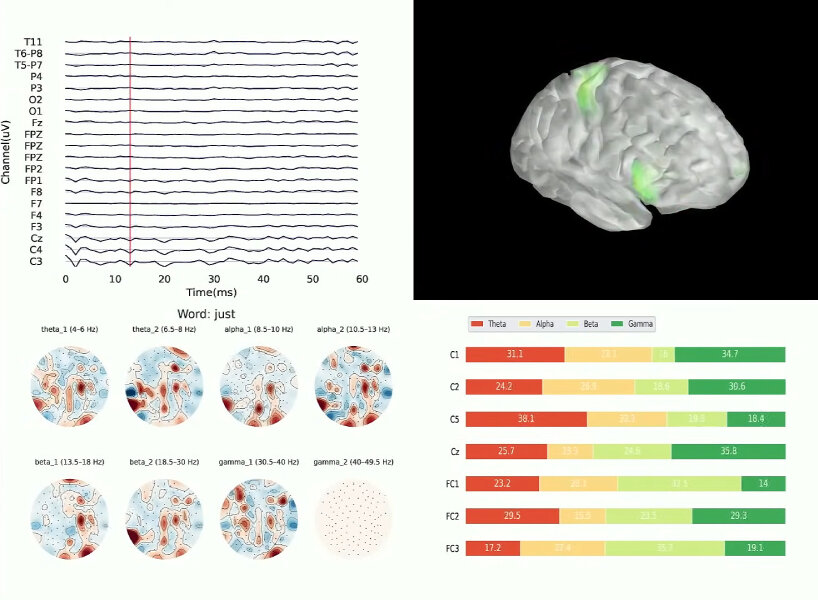}
\caption{T15 dataset overview showing iEEG waveforms, cortical topography maps for theta, alpha, beta, and gamma frequency bands, and electrode placement on speech motor cortex regions.}
\label{fig:dataset_overview}
\end{figure}

The 512-dimensional input feature vector encodes neural activity from four distinct brain regions, each measured via two complementary signal types, as detailed in Table~\ref{tab:channel_mapping}. Channels 0--256 record threshold crossings---discrete neural firing events detected when the extracellular voltage exceeds a calibrated amplitude threshold, providing a measure of single-unit and multi-unit spiking activity. Channels 257--512 record spike-band power---the continuous root-mean-square amplitude in the 300--1000~Hz band, capturing population-level neural activity that reflects aggregate firing rates across many neurons. Together, these dual measurements provide complementary spatiotemporal views of the same cortical regions: threshold crossings offer high temporal precision for individual spike events, while spike-band power provides a smoother, more robust signal less susceptible to electrode drift~\cite{willett2023speech}.

\begin{table}[!t]
\centering
\caption{Electrode Channel Mapping: 512 Input Features}
\label{tab:channel_mapping}
\begin{tabular}{@{}cll@{}}
\toprule
\textbf{Range} & \textbf{Brain Area} & \textbf{Measurement} \\
\midrule
0--64 & Ventral 6v & Threshold crossings \\
65--128 & Area 4 (M1) & Threshold crossings \\
129--192 & 55b & Threshold crossings \\
193--256 & Dorsal 6v & Threshold crossings \\
\midrule
257--320 & Ventral 6v & Spike band power \\
321--384 & Area 4 (M1) & Spike band power \\
385--448 & 55b & Spike band power \\
449--512 & Dorsal 6v & Spike band power \\
\bottomrule
\end{tabular}
\end{table}

\begin{figure}[!t]
\centering
\includegraphics[width=0.95\linewidth]{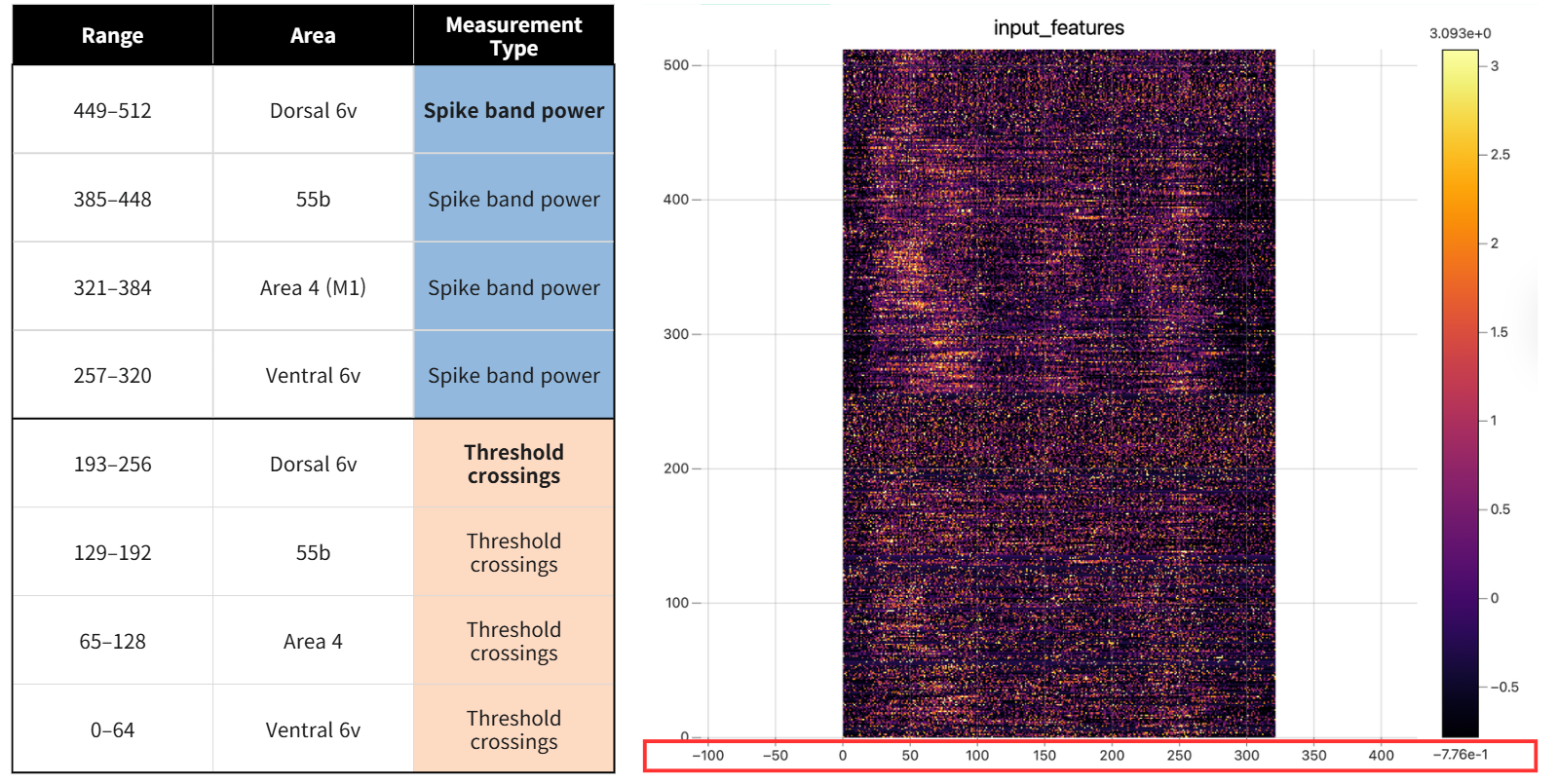}
\caption{Input feature heatmap for a representative trial. The y-axis shows 512 channels organized by brain region and measurement type; the x-axis shows time steps. The upper half (channels 0--256) shows threshold crossings; the lower half (257--512) shows spike-band power.}
\label{fig:input_heatmap}
\end{figure}

Each trial is stored in HDF5 format containing three fields: \texttt{input\_features} ($\in \mathbb{R}^{T \times 512}$), \texttt{seq\_class\_ids} (ground-truth phoneme sequence as integer indices into a 42-class vocabulary: 41 ARPAbet phonemes from the CMU Pronouncing Dictionary~\cite{cmudict2014} plus one CTC blank token at index 0), and \texttt{transcription} (the ASCII-encoded sentence stored as integer character codes, e.g., [87, 104, 97, 116, ...] for ``What...''). For example, trial\_0002 contains the sentence ``What do they like?'' encoded as phoneme sequence \texttt{[W AH T SIL D UW AO DH EY SIL L AY K SIL]} where SIL denotes silence boundaries between words. The average trial contains approximately 150 time steps (3 seconds at 50~Hz) and 12--15 phonemes, though this varies considerably: short trials such as ``Do it'' contain only 4 phonemes in approximately 50 time steps, while complex sentences such as ``I tell you my family is not hungry'' span 30+ phonemes across 400+ time steps. This variable-length characteristic necessitates the use of CTC~\cite{graves2006connectionist} rather than fixed-output architectures, as the input-output length ratio varies from approximately 3:1 to 15:1 across trials. The phoneme class distribution is highly non-uniform: the most frequent phoneme AH (the schwa, as in ``the'') constitutes approximately 10.3\% of all phoneme tokens in the training set, while rare phonemes such as ZH (as in ``vision'': 0.04\%) and OY (as in ``boy'': 0.11\%) appear in fewer than 0.2\% of tokens.

A critical characteristic of this dataset is the temporal progression in vocabulary difficulty. Early sessions (August--December 2023) contain simple conversational sentences such as ``Bring it closer,'' ``What do they like?'' and ``Do it,'' while sessions from 2024 introduce medium-complexity vocabulary, and 2025 sessions contain challenging domain-specific terms, brand names, and proper nouns (e.g., ``Barometric volume''). This creates a natural distribution shift that, combined with a 7-month recording gap between the 2024 and 2025 data, poses a significant generalization challenge.

\begin{figure}[!t]
\centering
\includegraphics[width=0.95\linewidth]{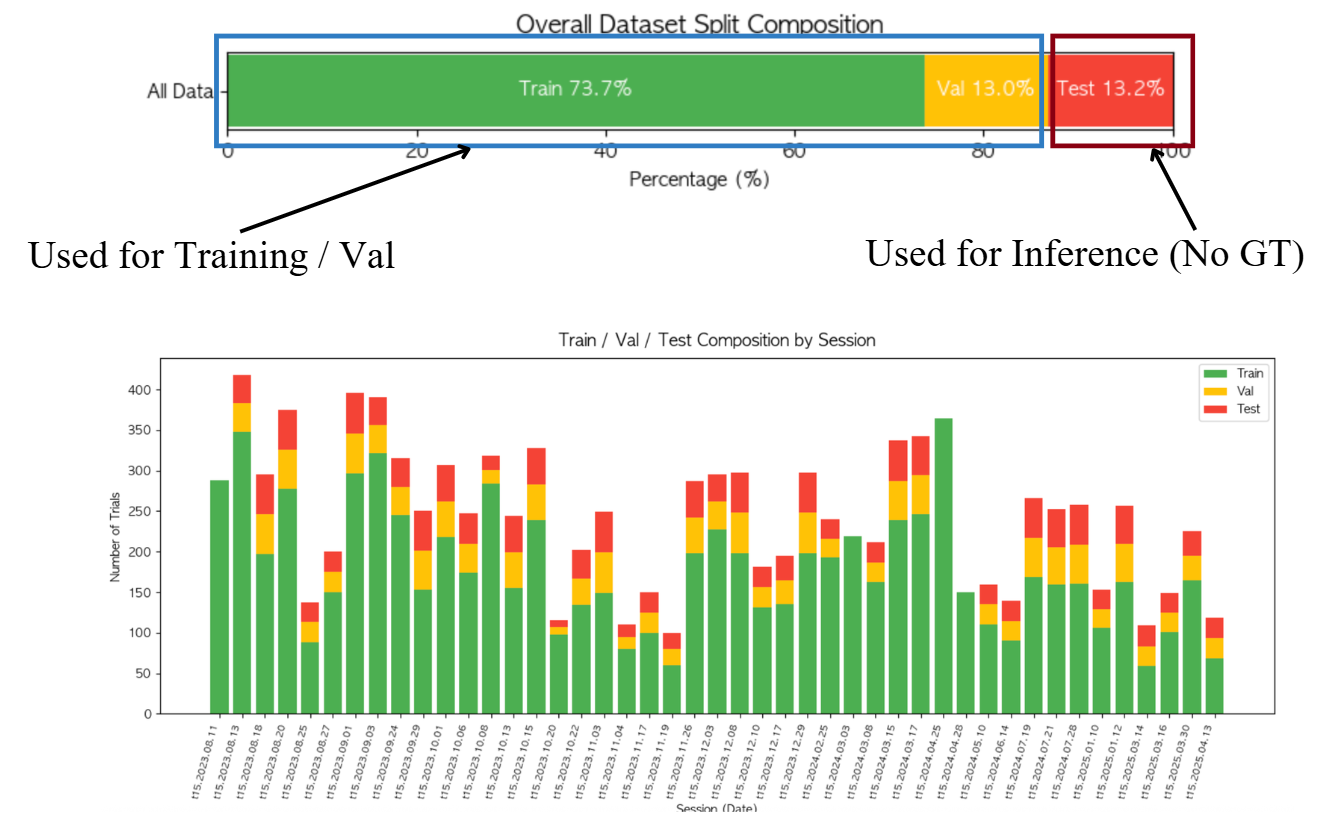}
\caption{Dataset split composition. Top: overall distribution. Bottom: per-session breakdown showing train (green), validation (yellow), and test (red) trials across all 45 sessions.}
\label{fig:data_split}
\end{figure}

The dataset is partitioned into training (73.7\%, 7,050 trials across 41 sessions), validation (13.0\%, 1,021 trials across 4 sessions spanning the full recording period), and test (13.2\%, used for inference only, no ground truth available). Fig.~\ref{fig:data_split} shows the per-session composition.

\section{Data Processing Pipeline}

The raw iEEG signals undergo a four-stage preprocessing pipeline that transforms the high-bandwidth, noisy 30~kHz recordings into clean, normalized feature matrices suitable for deep learning. The first stage applies a 4th-order Butterworth bandpass filter with cutoff frequencies $[0.3, 300]$~Hz, implemented via \texttt{scipy.signal.butter(N=4, Wn=[0.3, 300])}, to remove DC drift below 0.3~Hz and high-frequency noise above 300~Hz while preserving the physiologically relevant frequency bands: theta (4--8~Hz, associated with memory encoding and attentional gating), alpha (8--13~Hz, reflecting cortical idle states and sensorimotor mu rhythms), beta (13--30~Hz, linked to motor planning and sustained cortical activation), and gamma (30--100~Hz, correlated with local cortical processing and phoneme-specific articulatory commands)~\cite{bouchard2013functional,lotte2018review}. The transfer function $H(s) = [1 + (s/\omega_c)^{2N}]^{-1}$ with $N=4$ provides a sharp 80~dB/decade rolloff that effectively eliminates power line interference (50/60~Hz harmonics fall outside the passband in the stop-band region) and high-frequency electromyographic (EMG) artifacts from facial muscles without distorting the neural signal content within the passband.

The second stage applies Common Average Reference (CAR) spatial filtering, subtracting the mean signal across all channels at each time point: $x_{\text{CAR}}[c, t] = x[c, t] - C^{-1}\sum_{c'=1}^{C} x[c', t]$ where $C = 512$. This removes artifacts shared across electrodes---such as volume-conducted electrical noise from muscles, heartbeat, and environmental electromagnetic interference---while preserving spatially localized neural signals that differ across cortical regions. CAR is preferred over bipolar referencing for high-density arrays because it does not require assumptions about electrode pair geometry~\cite{lotte2018review}.

The third stage compresses the filtered signal from 30~kHz to 50~Hz through temporal binning, averaging within non-overlapping 20~ms windows: $x_{\text{bin}}[t] = 600^{-1}\sum_{k=600t}^{600(t+1)-1} x_{\text{CAR}}[k]$. This 600$\times$ compression reduces a 3-second trial from $[512, 90{,}000]$ to $[512, 150]$ time steps. The 20~ms bin width was chosen to match the approximate duration of the fastest phoneme transitions ($\sim$40~ms for stop consonants~\cite{jurafsky2009speech}), ensuring that at least two time bins capture each phoneme event. The averaging also serves as a low-pass anti-aliasing filter and improves signal-to-noise ratio by $\sqrt{600} \approx 24.5\times$.

The fourth stage applies per-session, per-channel z-score normalization: $z[c] = (x[c] - \mu_c) / \sigma_c$, where $\mu_c$ and $\sigma_c$ are computed across all time steps within a single recording session. Since electrode impedance drifts over the 20-month recording span due to tissue encapsulation, glial scarring, and electrochemical degradation~\cite{willett2023speech}, this may cause baseline shifts that would otherwise dominate the learned representations. The final output is a matrix $\in \mathbb{R}^{T \times 512}$ ready for batched model input as $\in \mathbb{R}^{B \times T \times 512}$.

\begin{figure*}[!t]
\centering
\includegraphics[width=0.95\linewidth]{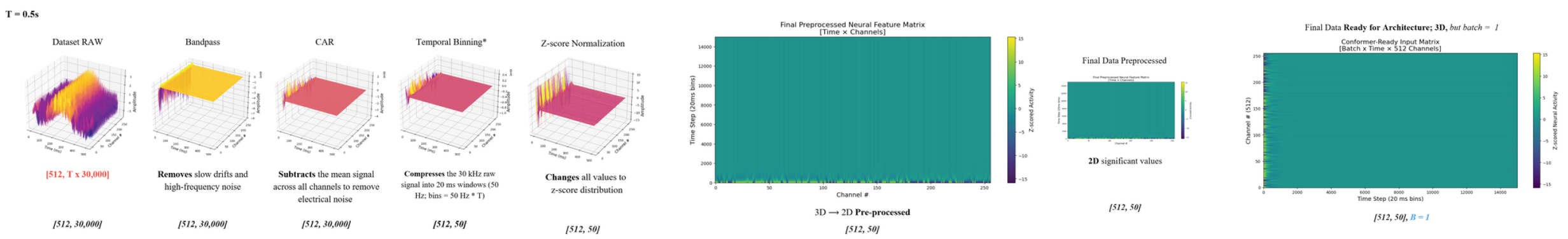}
\caption{Data processing pipeline. 3D surface plots show the 512-channel signal at each stage for a 0.5s trial (T=0.5s). Bandpass filtering removes drift and HF noise; CAR subtracts common-mode artifacts; temporal binning compresses 600$\times$; z-score normalization standardizes across sessions.}
\label{fig:preprocessing}
\end{figure*}

\section{Model Architecture}

\begin{table*}[!t]
\centering
\caption{Architectural Comparison: Standard Conformer vs. ConformerXL. Components marked ``Novel'' are new to this work; ``Modified'' indicates adapted components with changed parameters.}
\label{tab:arch_comparison}
\begin{tabular}{@{}llllclcc@{}}
\toprule
\textbf{Component} & \textbf{Standard Conformer} & \textbf{ConformerXL (Ours)} & \textbf{Status} & \quad & \textbf{Parameter} & \textbf{Standard} & \textbf{Ours} \\
\midrule
\multicolumn{4}{l}{\textit{Prenet}} & & d\_model & 512 & \textbf{384} \\
Structure & 2$\times$ Conv2D + Linear & Dilated Conv + BiGRU & Novel & & num\_layers & 12--17 & \textbf{12} \\
Convolutions & Single scale & Multi-scale (d=1, d=2) & Novel & & num\_heads & 8 & \textbf{6} \\
Temporal modeling & None & Bidirectional GRU & Novel & & head\_dim & 64 & 64 \\
Kernel size & 3 & 5 & Novel & & ff\_expansion & 4 & 4 \\
Normalization & LayerNorm & RMSNorm & Novel & & conv\_kernel & 31 & \textbf{15} \\
\midrule
\multicolumn{4}{l}{\textit{Subsampling}} & & dropout & 0.1 & \textbf{0.15} \\
Factor & 4$\times$ (2 layers) & 8$\times$ (3 layers) & Novel & & & & \\
Conv type & Conv2D & Conv1D & Novel & & \multicolumn{3}{l}{\textbf{System Specifications}} \\
Activation & ReLU & GELU & Novel & & Total params & --- & \textbf{192.9M} \\
\midrule
\multicolumn{4}{l}{\textit{Conformer Block}} & & Model size & --- & \textbf{1.47 GB} \\
Norm type & LayerNorm & RMSNorm & Modified & & Encoder layers & --- & \textbf{12} \\
Norm position & Post-norm & Pre-norm & Modified & & Hidden dim & --- & \textbf{384} \\
Conv module norm & BatchNorm & GroupNorm(32) & Modified & & FFN dim & --- & \textbf{1536} \\
Conv kernel size & 31 & 15 & Modified & & ConformerXL & --- & $\sim$25M \\
Block order & FFN$\to$MHSA$\to$Conv$\to$FFN & Same & Same & & WFST decoder & --- & 0 (algo) \\
Residual weights & 0.5, 1.0, 1.0, 0.5 & Same & Same & & 6-gram LM & --- & 3.1M entries \\
\midrule
\multicolumn{4}{l}{\textit{Output}} & & Total system & --- & $\sim$2.2 GB \\
Final norm & Inside last block & Separate RMSNorm & Modified \\
\bottomrule
\end{tabular}
\end{table*}

\begin{figure}[!t]
\centering
\includegraphics[width=0.95\linewidth]{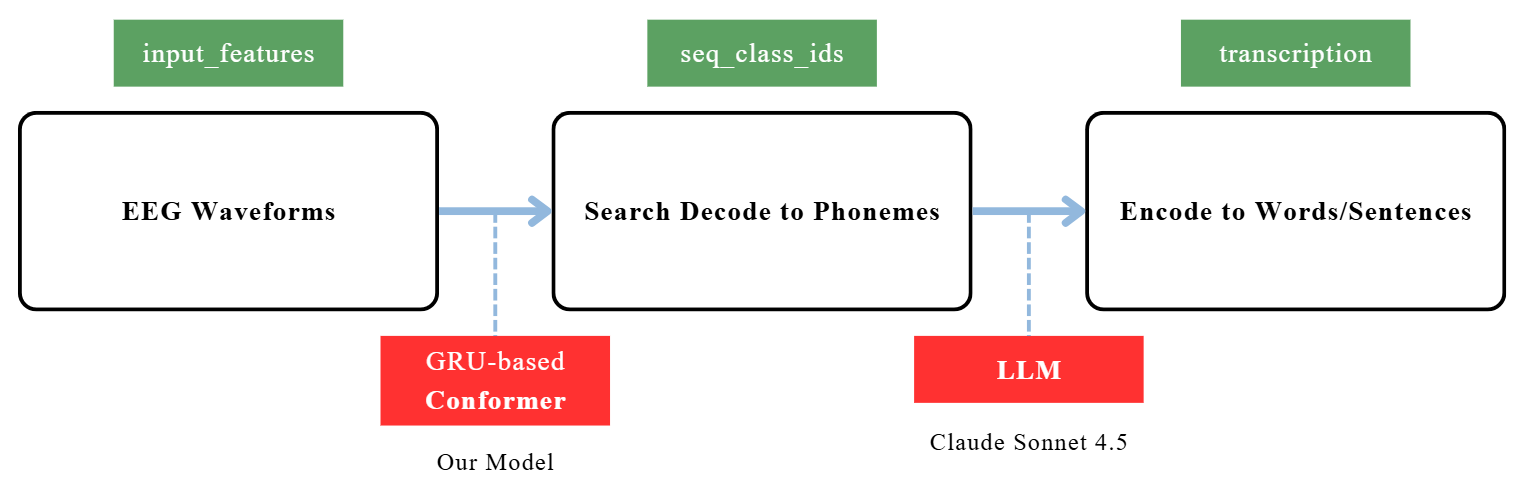}
\caption{End-to-end pipeline. The ConformerXL decodes iEEG to phonemes; an LLM encodes phonemes to sentences.}
\label{fig:pipeline}
\end{figure}

The ConformerXL is a 192.9M-parameter acoustic model that maps 512-channel iEEG input to 42-class phoneme output through three processing stages: a temporal prenet for multi-scale feature extraction and neural jitter correction, 8$\times$ temporal subsampling for CTC-compatible sequence compression, and 12 Conformer encoder blocks for global-local feature integration. The decoded phoneme sequences are subsequently converted to natural language sentences using Claude Sonnet 4.5 as an LLM-based phoneme-to-word encoder. Fig.~\ref{fig:pipeline} shows the end-to-end data flow, and Table~\ref{tab:arch_comparison} provides a comprehensive comparison with the standard Conformer architecture.

\begin{figure*}[!t]
\centering
\includegraphics[width=0.75\linewidth]{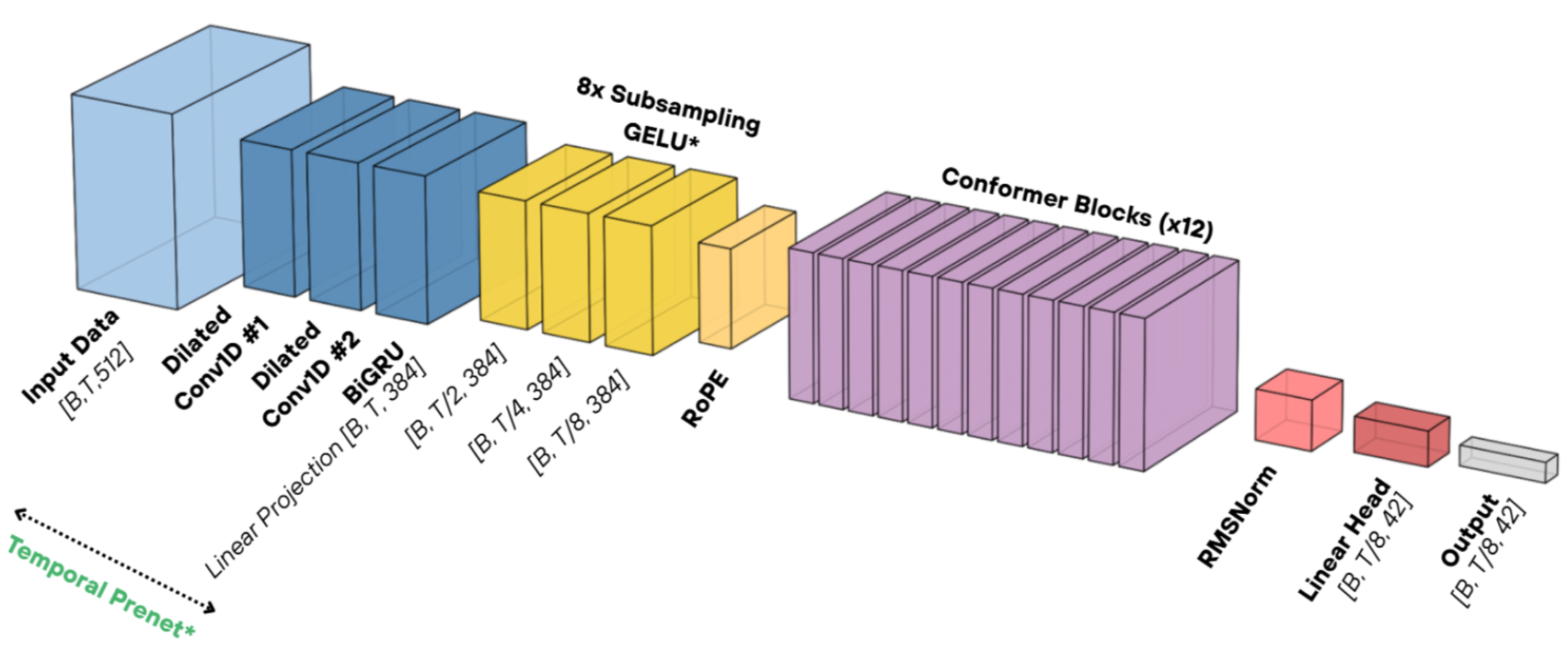}
\caption{ConformerXL architecture. The temporal prenet extracts multi-scale features via dilated convolutions and BiGRU before 8$\times$ subsampling compresses the sequence. Twelve Pre-RMSNorm Conformer blocks process encoded features, followed by linear projection to 42 phoneme classes.}
\label{fig:architecture}
\end{figure*}

The temporal prenet is our primary architectural contribution, motivated by the fundamental differences between audio speech signals and intracranial neural recordings. Audio features (e.g., 80-dimensional log-mel spectrograms) are generated directly by the articulatory process being decoded and exhibit relatively clean temporal structure. In contrast, iEEG recordings contain substantial inter-channel noise from volume conduction, electrode-specific impedance artifacts, and neural signals operating on multiple temporal scales simultaneously---fast spike events ($<$5~ms), medium-timescale local field potentials (10--100~ms), and slow cortical state fluctuations ($>$100~ms)~\cite{bouchard2013functional}. Standard Conformer prenets (two Conv2D layers with a linear projection) process all temporal scales uniformly, losing the multi-scale structure that is critical for distinguishing phonemically relevant neural patterns from noise.

Our prenet addresses this through two dilated 1D convolutions followed by a bidirectional GRU and linear projection:
\begin{align}
    h_1 &= \text{ReLU}(\text{RMSNorm}(\text{Conv1D}(X, k{=}5, d{=}1))) \\
    h_2 &= \text{ReLU}(\text{RMSNorm}(\text{Conv1D}(h_1, k{=}5, d{=}2))) \\
    h_3 &= \text{BiGRU}(h_2, \text{hidden}{=}256) \\
    X_{\text{prenet}} &= \text{Linear}(h_3) \in \mathbb{R}^{B \times T \times 384}
\end{align}
The dilation=1 convolution (effective receptive field: 5 time steps = 100~ms at 50~Hz) captures fast temporal patterns such as individual spike events and rapid phoneme transitions, while the dilation=2 convolution (effective receptive field: 9 time steps = 180~ms) captures slower dynamics such as sustained articulatory postures and phoneme-level temporal envelopes. This multi-scale decomposition is analogous to wavelet-based approaches in neural signal processing~\cite{lotte2018review} but is learned end-to-end rather than requiring hand-designed filter banks.

\begin{figure}[!t]
\centering
\includegraphics[width=0.95\linewidth]{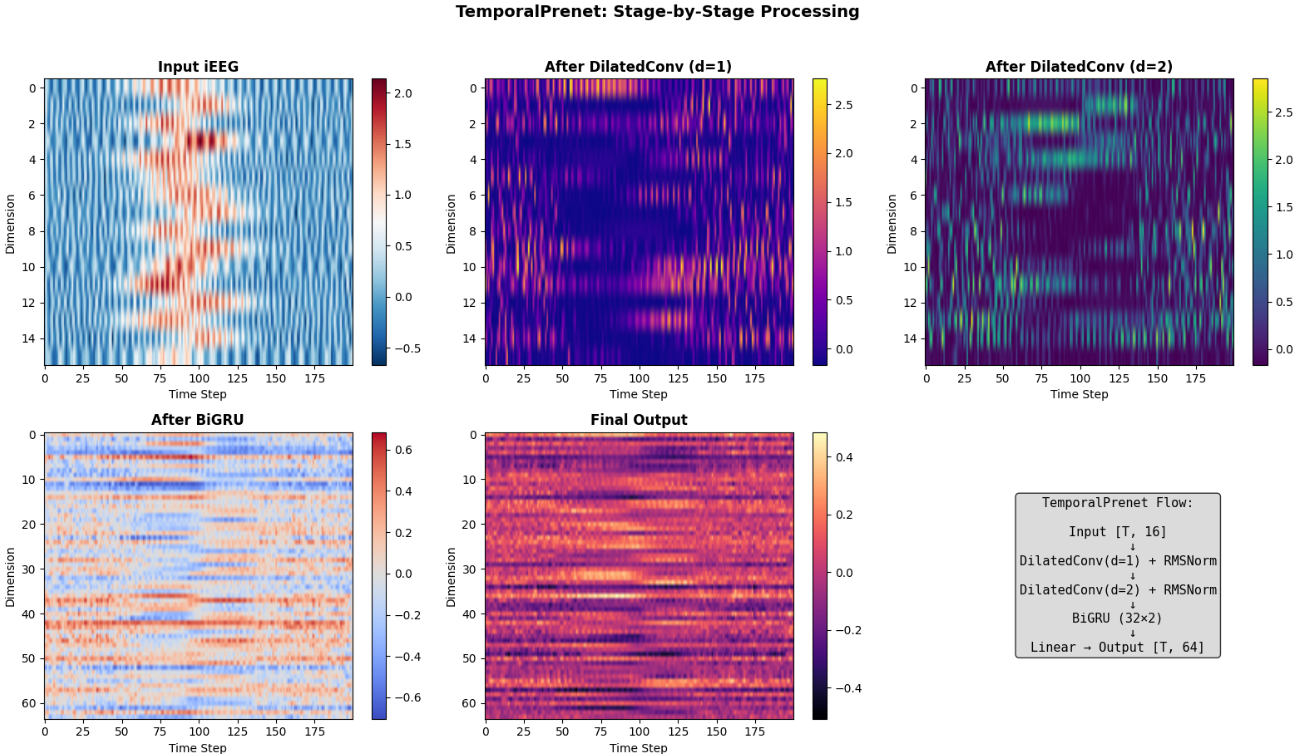}
\caption{Temporal prenet stage-by-stage processing. Dilated convolution d=1 captures fast transitions; d=2 captures sustained patterns; BiGRU fuses bidirectional context; linear projection maps to model dimension.}
\label{fig:temporal_prenet}
\end{figure}

\begin{figure}[!t]
\centering
\includegraphics[width=0.95\linewidth]{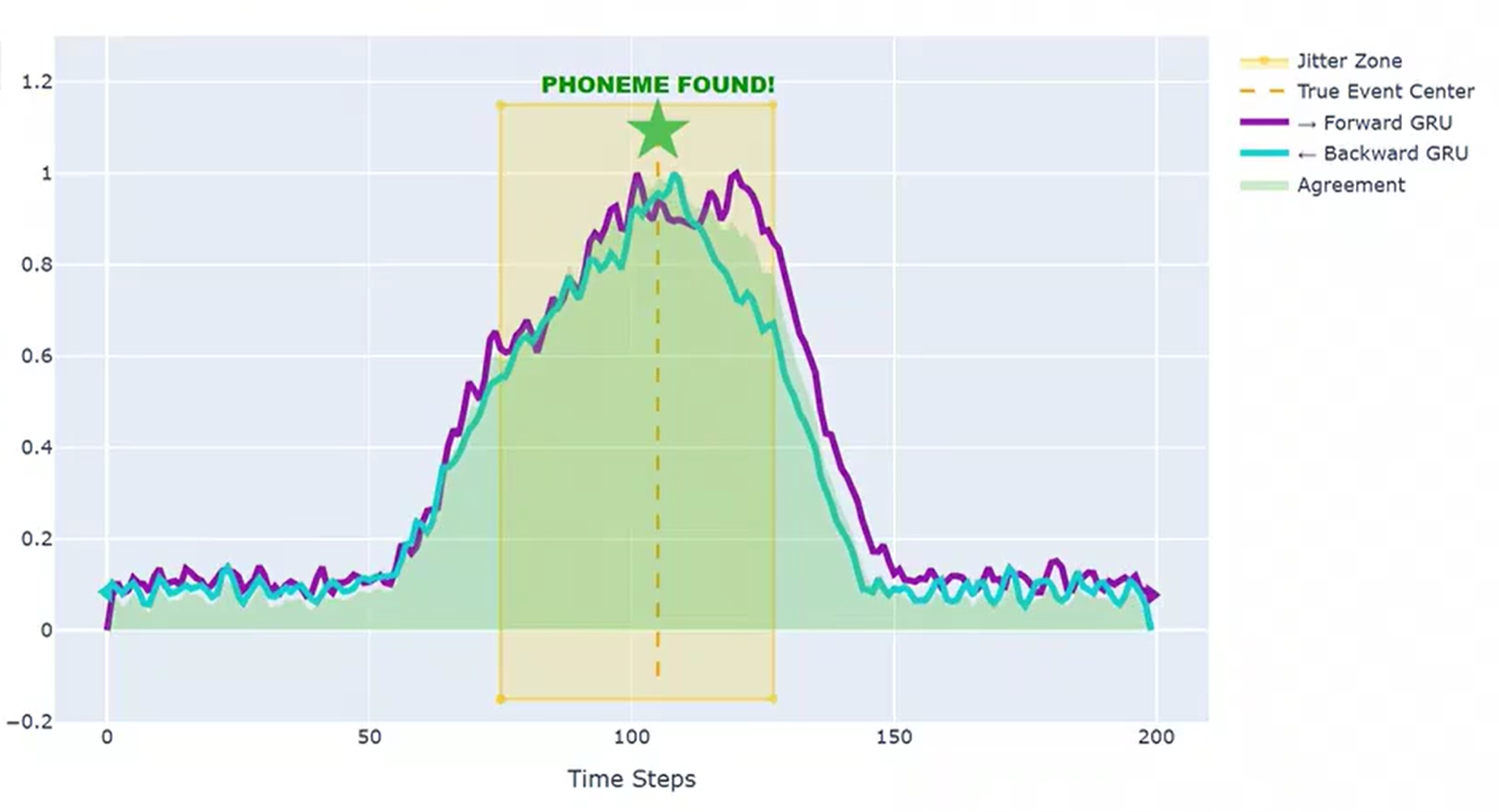}
\caption{BiGRU temporal jitter correction. Forward and backward GRU outputs converge at the true phoneme event center, correcting for 10--50~ms timing variability in neural signals.}
\label{fig:bigru_jitter}
\end{figure}

The bidirectional GRU~\cite{cho2014learning} following the dilated convolutions corrects for temporal jitter in neural signals. The GRU computes, at each time step $t$, an update gate $z_t$, reset gate $r_t$, and candidate hidden state $\tilde{h}_t$:
\begin{align}
    z_t &= \sigma(W_z x_t + U_z h_{t-1} + b_z) \\
    r_t &= \sigma(W_r x_t + U_r h_{t-1} + b_r) \\
    \tilde{h}_t &= \tanh(W_h x_t + U_h (r_t \odot h_{t-1}) + b_h) \\
    h_t &= (1 - z_t) \odot h_{t-1} + z_t \odot \tilde{h}_t
\end{align}
where $\sigma$ is the sigmoid function and $\odot$ denotes element-wise multiplication. The update gate $z_t$ controls how much of the new candidate state to incorporate versus retaining the previous state, which is particularly relevant for neural signals where the same phoneme may be represented by activity shifted by 10--50~ms between trials due to variability in motor planning latency, neural conduction delays, and electrode coupling dynamics. In the bidirectional configuration, the forward GRU $\overrightarrow{h}_t$ processes left-to-right while the backward GRU $\overleftarrow{h}_t$ processes right-to-left, and their outputs are concatenated: $h_t = [\overrightarrow{h}_t; \overleftarrow{h}_t] \in \mathbb{R}^{512}$ (256 hidden units per direction). The linear projection then maps this to $\mathbb{R}^{384}$. By processing each time step with context from both past and future, the BiGRU effectively aligns the neural representation to a canonical temporal frame where the forward and backward hidden states converge or meet at the true phoneme event center, regardless of trial-to-trial timing variability. This convergence behavior seems to be robust because the GRU's gating mechanism can learn to ``wait'' for confirmatory evidence from the opposite direction before committing to a phoneme boundary decision. Fig.~\ref{fig:bigru_jitter} visualizes this alignment process.

The 8$\times$ temporal subsampling tries to address a constraint of CTC training: the subsampled input length $T'$ must exceed the target phoneme sequence length $L$, or the CTC loss becomes undefined. For iEEG at 50~Hz, raw trial lengths range from 700--1,400 frames while target sequences contain only 20--45 phonemes. With standard 4$\times$ subsampling, the ratio $T'/L$ remains excessively large ($\sim$7--15$\times$), causing CTC to assign most frames to blank tokens, which produces sparse gradient signals and training instability. Our 8$\times$ subsampling uses three consecutive stride-2 Conv1D layers with GELU activation:
\begin{equation}
    X_{\text{sub}} = f_3(f_2(f_1(X))), \quad f_i(x) = \text{GELU}(\text{Conv}_{s=2}^{(i)}(x))
\end{equation}
This reduces $T'/L$ to a more balanced $\sim$2--4$\times$ range, substantially improving CTC stability and convergence speed. The choice of GELU~\cite{hendrycks2016gaussian} over ReLU is needed. GELU preserves subtle signal variations near zero that ReLU would eliminate, which matters for neural signals where small amplitude differences between phonemically relevant and irrelevant activity carry discriminative information.

Each of the 12 ConformerXL encoder blocks follows the macaron-style sandwich with Pre-RMSNorm~\cite{zhang2019root} applied before each sub-layer. The computation within a single block proceeds as:
\begin{align}
    \tilde{x}^{(1)} &= x + \tfrac{1}{2}\text{FFN}_1(\text{RMSNorm}(x)) \\
    \tilde{x}^{(2)} &= \tilde{x}^{(1)} + \text{MHSA}(\text{RMSNorm}(\tilde{x}^{(1)})) \\
    \tilde{x}^{(3)} &= \tilde{x}^{(2)} + \text{ConvModule}(\text{RMSNorm}(\tilde{x}^{(2)})) \\
    y &= \tilde{x}^{(3)} + \tfrac{1}{2}\text{FFN}_2(\text{RMSNorm}(\tilde{x}^{(3)}))
\end{align}
The half-step residual weighting (0.5) on the FFN sub-layers and full-step weighting (1.0) on MHSA and convolution follows the Macaron-Net finding~\cite{gulati2020conformer} that sandwiching attention between two half-step feed-forward layers improves gradient flow and representation capacity compared to a single full-step FFN.

The feed-forward module applies a two-layer MLP with GELU~\cite{hendrycks2016gaussian} activation and 4$\times$ expansion:
\begin{equation}
    \text{FFN}(x) = W_2 \cdot \text{GELU}(W_1 x + b_1) + b_2
\end{equation}
where $W_1 \in \mathbb{R}^{384 \times 1536}$ expands the representation to capture richer feature interactions, and $W_2 \in \mathbb{R}^{1536 \times 384}$ projects back. Dropout of 0.15 is applied after each linear layer. The GELU activation $\text{GELU}(x) = x \cdot \Phi(x)$, where $\Phi$ is the Gaussian CDF, provides a smooth approximation to ReLU that preserves gradient information for near-zero inputs---particularly important given the low-amplitude neural signals that distinguish phonemically relevant from irrelevant cortical activity.

The multi-head self-attention module computes scaled dot-product attention across 6 heads with head dimension $d_k = 64$:
\begin{equation}
    \text{Attention}(Q, K, V) = \text{Softmax}\left(\frac{QK^T}{\sqrt{d_k}}\right)V
\end{equation}
where $Q = xW^Q$, $K = xW^K$, $V = xW^V$ with $W^Q, W^K, W^V \in \mathbb{R}^{384 \times 64}$ per head. The six heads independently attend to different temporal relationships in the subsampled iEEG sequence ($T/8$ time steps), capturing dependencies ranging from adjacent-phoneme transitions to sentence-level prosodic patterns. Sinusoidal positional encodings inject absolute position information, since the self-attention mechanism is otherwise permutation-invariant and would lose temporal ordering.

The convolution module applies, in sequence: (1) a pointwise convolution expanding channels from 384 to 768 ($2\times$), (2) a Gated Linear Unit (GLU) that splits the 768 channels into two halves and applies sigmoid gating---$\text{GLU}(x) = x_a \otimes \sigma(x_b)$ where $x_a, x_b \in \mathbb{R}^{384}$---halving the channels back to 384, (3) a depthwise separable convolution with kernel size 15 that processes each channel independently, (4) GroupNorm with 32 groups for within-batch normalization, (5) Swish activation $\text{Swish}(x) = x \cdot \sigma(x)$, and (6) a final pointwise convolution projecting back to 384 dimensions. The depthwise convolution kernel size of 15 (vs. 31 in the standard Conformer) was selected because the 8$\times$ subsampled iEEG at 50~Hz has effective temporal resolution of $8 \times 20\text{ms} = 160$~ms per frame, meaning a kernel of 15 spans $15 \times 160 = 2.4$~seconds---sufficient to capture the longest phoneme durations in English~\cite{jurafsky2009speech} without introducing unnecessary computation from oversized kernels. The replacement of BatchNorm with GroupNorm(32) addresses a practical issue: with batch sizes of 16 and variable-length sequences, BatchNorm statistics are unreliable, while GroupNorm computes statistics within groups of channels independently of batch size.

The replacement of all LayerNorm with Root Mean Square Normalization~\cite{zhang2019root} across the entire architecture was done, as training with standard LayerNorm produced NaN losses at approximately epoch 50--70 when the learning rate was still relatively high under cosine scheduling. RMSNorm normalizes by the root mean square without re-centering:
\begin{equation}
    \text{RMSNorm}(x) = \frac{x}{\sqrt{\frac{1}{d}\sum_{i=1}^{d}x_i^2 + \epsilon}} \cdot \gamma
\end{equation}
where $\gamma$ is a learnable scale parameter and $\epsilon = 10^{-8}$ prevents division by zero. The omission of mean subtraction (present in LayerNorm) means RMSNorm does not assume zero-mean inputs, which is more appropriate for neural signals whose mean shifts across sessions, electrodes, and even within trials as cortical state fluctuates. RMSNorm also has a computational advantage. It requires one fewer reduction operation per normalization, which accumulates across the 48 RMSNorm instances (4 per block $\times$ 12 blocks) evaluated at every training step.

\begin{figure*}[!t]
\centering
\includegraphics[width=0.8\linewidth]{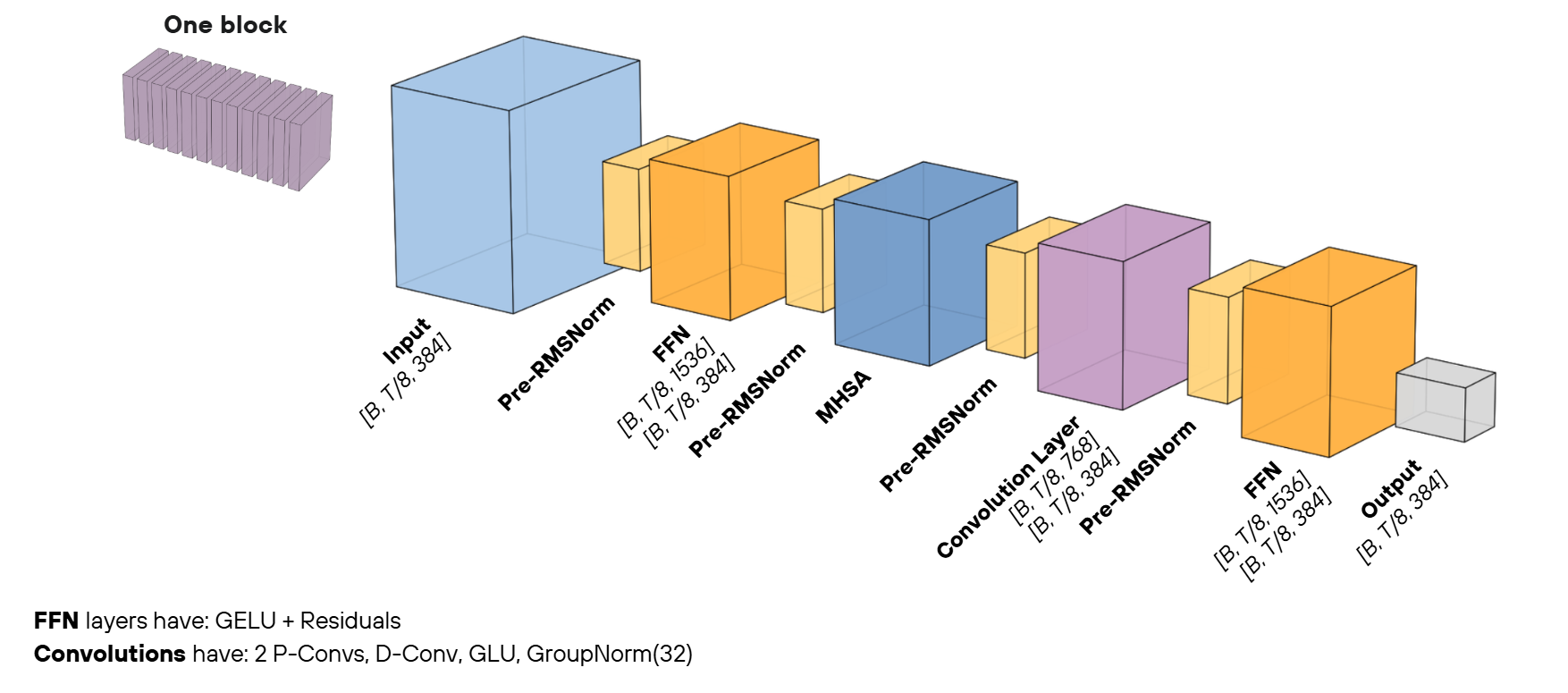}
\caption{Single ConformerXL block with Pre-RMSNorm before each sub-layer. FFN uses GELU + 0.5$\times$ residual; convolution uses pointwise-depthwise-pointwise with GLU and GroupNorm(32).}
\label{fig:conformer_block}
\end{figure*}

After the 12 encoder blocks, a separate RMSNorm followed by a linear projection maps to 42 output classes: $P(\text{phoneme}|\text{frame}) = \text{Softmax}(\text{Linear}(\text{RMSNorm}(h_{12})))$.

\section{Training Methodology}

\begin{figure}[!t]
\centering
\includegraphics[width=0.95\linewidth]{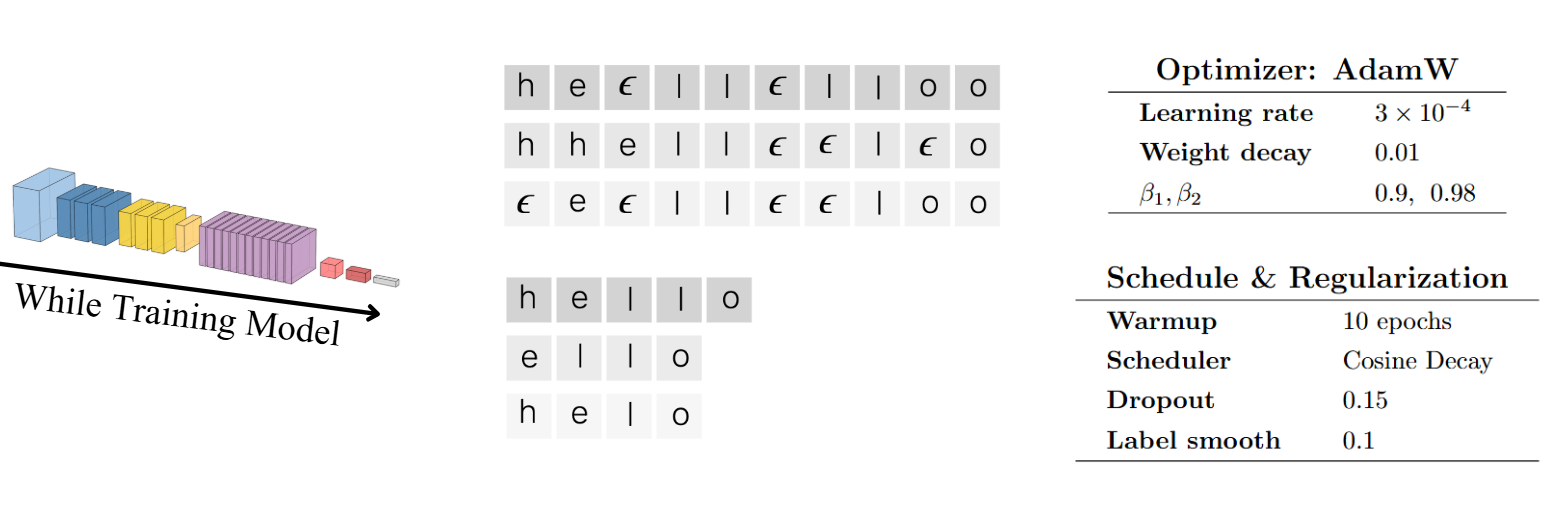}
\caption{CTC training. Multiple alignment paths through blank and phoneme tokens collapse to the same output. Right: optimizer and regularization configuration.}
\label{fig:ctc}
\end{figure}

We train with CTC loss~\cite{graves2006connectionist}, which marginalizes over all valid alignment paths that collapse to the target phoneme sequence. Given input $X$ and target $y$, the loss is:
\begin{equation}
    \mathcal{L}_{\text{CTC}} = -\log \sum_{\pi \in \mathcal{B}^{-1}(y)} P(\pi \mid X)
\end{equation}
where $\mathcal{B}$ is the CTC collapse operator that merges repeated tokens and removes blanks. The forward-backward algorithm~\cite{graves2006connectionist} computes this marginalization efficiently by defining forward variables $\alpha(t,s)$ and backward variables $\beta(t,s)$ over a lattice of $T$ time steps and $U = |y|$ target symbols, yielding the total probability as:
\begin{equation}
    P(y \mid X) = \sum_s \alpha(T, s) \cdot \beta(T, s)
\end{equation}
This runs in $O(T \times U)$ time, avoiding the exponential cost of enumerating all $42^T$ possible alignments. For a typical trial with $T/8 = 19$ subsampled frames and $U = 12$ phonemes, the lattice requires only about 228 cell evaluations rather than $42^{19} \approx 10^{30}$ path enumerations. Fig.~\ref{fig:ctc} illustrates how multiple paths through blank and phoneme tokens collapse to the same output.

We use AdamW~\cite{loshchilov2019decoupled} with learning rate $3 \times 10^{-4}$, decoupled weight decay 0.01, and betas $(0.9, 0.98)$. The $\beta_2 = 0.98$ (rather than the default 0.999) was chosen to provide faster adaptation of the second moment estimates, which is beneficial when training on non-stationary neural data where gradient statistics can shift between sessions. The learning rate schedule consists of 10-epoch linear warmup---during which the rate increases from 0 to $3 \times 10^{-4}$---followed by cosine decay: $\eta_t = \eta_{\min} + \frac{1}{2}(\eta_{\max} - \eta_{\min})(1 + \cos(\pi t / T_{\max}))$, which provides smooth annealing without the abrupt transitions of step-wise schedules.

Regularization is applied at multiple levels of the architecture. Dropout of 0.15 is applied after every linear transformation and attention computation throughout the network, which is higher than the standard Conformer's 0.1 to compensate for the smaller training set (7,050 trials vs. the 960 hours typical of ASR training). Label smoothing of 0.1 replaces hard one-hot targets with $P(y_{\text{correct}}) = 0.9$ and $P(y_{\text{other}}) = 0.1/41$ for each non-target class, preventing the model from becoming overconfident on training examples and improving generalization to unseen phoneme contexts. Gradient clipping at max norm 1.0 prevents the exploding gradients that can occur when CTC loss spikes on difficult alignment examples. Additionally, we adapt SpecAugment~\cite{park2019specaugment}---originally designed for mel-spectrogram masking in ASR---for iEEG by applying 3 time masks of up to 100 frames each (randomly zeroing contiguous time spans) and 2 channel masks of up to 25 channels each (randomly zeroing contiguous electrode groups). The channel masking forces the model to decode from partial electrode subsets, improving robustness to individual electrode failures or impedance degradation that can occur in chronic implants~\cite{willett2023speech}.

Training uses PyTorch~\cite{paszke2019pytorch} mixed precision (FP16 activations, FP32 weights and gradient accumulation) on a single GPU with batch size 16 and gradient accumulation of 2 (effective batch 32), completing in approximately 4 hours per run. The mixed precision scheme reduces memory consumption by approximately 40\% and training time by approximately 25\%, with negligible accuracy impact due to the FP32 master weights and loss scaling. Hyperparameter optimization uses Optuna~\cite{akiba2019optuna} with Tree-structured Parzen Estimator (TPE) sampling tracked via Weights \& Biases; decoder parameters were specifically swept across 150 trials over beam size $\in [32, 256]$, LM weight $\in [0.5, 1.5]$, and length penalty $\in [0.7, 1.2]$.

\section{Decoding Pipeline}

The decoding pipeline progresses through three stages. The greedy baseline (89.38\% accuracy, 45~ms latency) selects the highest-probability phoneme at each frame via $\hat{y}_t = \arg\max_c P(c | \text{frame}_t)$ and applies CTC collapse rules. This approach uses no contextual information whatsoever---each frame is decoded independently, which means phonotactically impossible sequences (e.g., five consecutive stop consonants) can occur in the output.

The second stage (90.02\%, 65~ms) integrates a 6-gram phoneme language model with modified Kneser-Ney smoothing~\cite{chen1999empirical} trained on 3.1M phoneme sequences from three sources: 600K sequences from the CMU Pronouncing Dictionary~\cite{cmudict2014} (providing coverage of standard English word-level phonotactics using the same 41-phoneme ARPAbet index as our model output), 2.0M sequences from LibriSpeech~\cite{panayotov2015librispeech} transcripts converted to phoneme sequences via grapheme-to-phoneme alignment (providing sentence-level phonotactic patterns including cross-word boundary transitions), and 7K sequences from the T15 training data (providing participant-specific articulatory patterns). The resulting LM has vocabulary size 42, perplexity 8.3, and size 1.7~MB.

The Kneser-Ney smoothing assigns probabilities to n-gram sequences by interpolating between the maximum likelihood estimate and a lower-order backoff distribution:
\begin{align}
    P_{\text{KN}}&(w_i | w_{i-n+1}^{i-1}) = \frac{\max(c(w_{i-n+1}^{i}) - D,\; 0)}{c(w_{i-n+1}^{i-1})} \nonumber \\
    &+ \gamma(w_{i-n+1}^{i-1}) \cdot P_{\text{KN}}(w_i | w_{i-n+2}^{i-1})
\end{align}
where $D$ is the discount parameter, $c(\cdot)$ denotes the count, and $\gamma$ is the backoff weight ensuring proper normalization. The recursive backoff is essential, in which, if a specific 6-gram context $p_{i-5:i-1}$ has not been observed in training, the model successively tries 5-gram, 4-gram, 3-gram, 2-gram, and unigram contexts until a match is found. Without this recursive mechanism, unseen 6-grams would receive a uniform fallback probability of $1/42 \approx 0.024$, which provides no discrimination between phonemically plausible and implausible continuations. In early experiments, we implemented a flat backoff (uniform probability for unseen n-grams) and observed that the LM-augmented decoder produced 88.2\% accuracy---\textit{worse} than the 89.38\% greedy baseline---because the uniform fallback dominated the scoring for novel phoneme contexts in the 2024--2025 sessions with complex vocabulary. Implementing full recursive Kneser-Ney backoff restored the LM's discriminative power, yielding the 90.02\% reported here.

The LM is combined with greedy output via log-linear interpolation with $\beta = 0.8$ weighting the LM heavily:
\begin{equation}
    S(p_i) = (1{-}\beta) \log P_{\text{AM}}(p_i) + \beta \log P_{\text{LM}}(p_i | p_{i-5:i-1})
\end{equation}
This heavy LM weighting reflects the observation that phonotactic constraints are more reliable than individual frame-level acoustic predictions for neural signals. A confusion matrix derived from systematic acoustic model errors on the validation set further informs the LM about which phoneme substitutions are most likely. For example, the model systematically confuses AO (as in ``bought'') with AA (as in ``bot'') and EY (as in ``say'') with EH (as in ``said''), both of which are articulatorily adjacent pairs that produce similar motor cortex activation patterns. The confusion-informed LM applies higher correction weights to these known confusion pairs.

\begin{figure}[!t]
\centering
\includegraphics[width=0.95\linewidth]{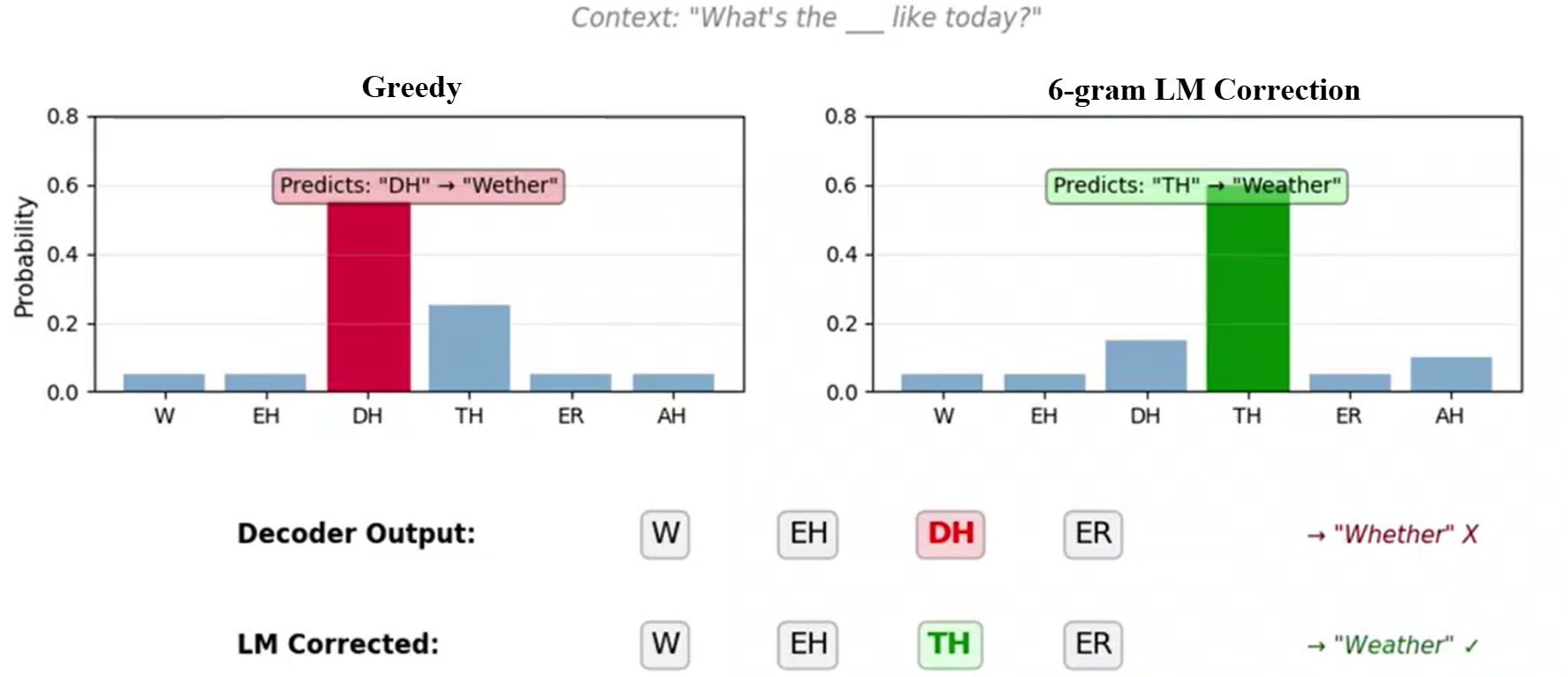}
\caption{Language model correction. Greedy decoding produces ``Wether''; the 6-gram LM applies phonotactic constraints to correct to ``Whether'' given preceding context.}
\label{fig:lm_correction}
\end{figure}

The third stage (92.14\%, 180~ms) replaces local rescoring with global search through a Weighted Finite State Transducer~\cite{mohri2002weighted} graph. Formally, a WFST is defined as a tuple:
\begin{equation}
    \mathcal{T} = (\Sigma,\; \Delta,\; Q,\; I,\; F,\; E,\; \lambda,\; \rho)
\end{equation}
where $\Sigma$ is the input alphabet (42 CTC output tokens), $\Delta$ is the output alphabet (41 phonemes), $Q$ is a finite set of states, $I \subseteq Q$ and $F \subseteq Q$ are the initial and final state sets, $E$ is the set of weighted transitions, and $\lambda: I \to \mathbb{R}$, $\rho: F \to \mathbb{R}$ assign initial and final weights~\cite{mohri2002weighted}. The weight of a path $\pi$ through the transducer is the product (in the log semiring, the sum) of all edge weights:
\begin{equation}
    w[\pi] = \lambda(q_0) \otimes \bigotimes_{i=1}^{|\pi|} w(e_i) \otimes \rho(q_{|\pi|})
\end{equation}
In our formulation, edge weights encode the 6-gram phoneme LM log-probabilities, so traversing the WFST effectively computes the joint acoustic-linguistic score for any phoneme sequence hypothesis.

\begin{figure}[H]
\centering
\includegraphics[width=0.95\linewidth]{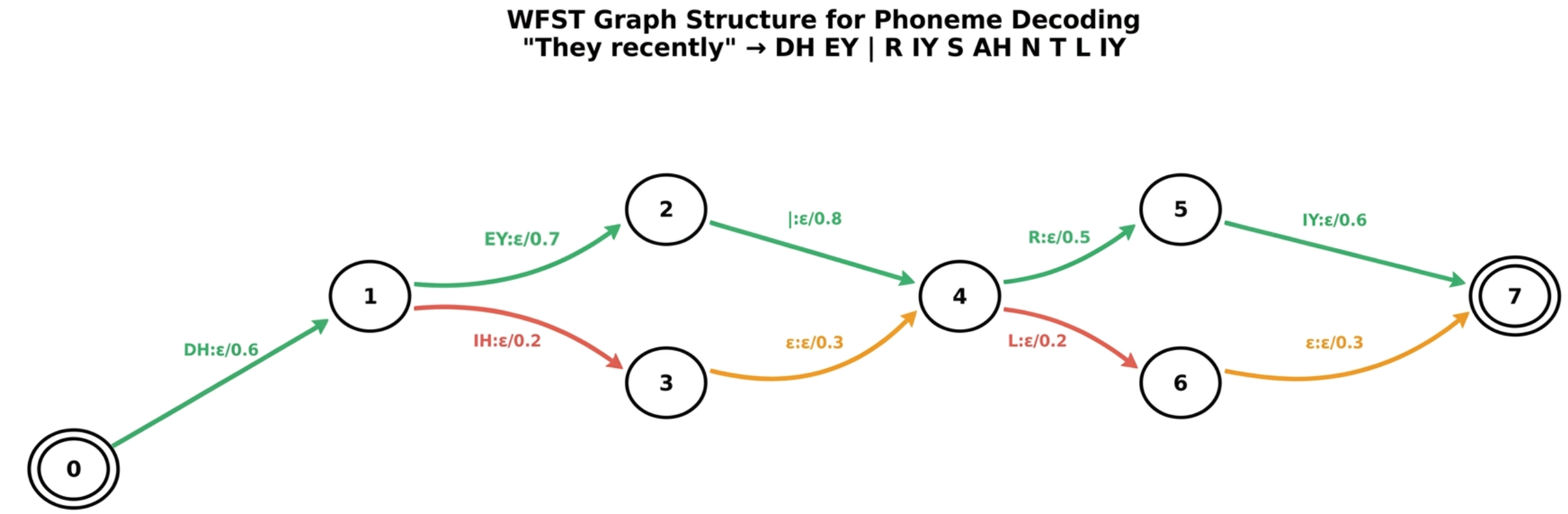}
\caption{WFST graph structure. Nodes are states; edges are LM-weighted phoneme transitions. Green: highest-scoring path.}
\label{fig:wfst_graph}
\end{figure}

The total score for a decoded sequence $y = (p_1, \ldots, p_n)$ given acoustic observations $\mathbf{x} = (x_1, \ldots, x_T)$ combines acoustic model evidence with the WFST path weight, normalized by sequence length:
\begin{align}
    \text{Score}(y) = \frac{1}{n^{\alpha}} \bigg( &\sum_{t=1}^{T}\log P_{\text{AM}}(p_t | \mathbf{x}_t) \nonumber \\
    &+ \lambda \sum_{i=1}^{n}\log P_{\text{LM}}(p_i | p_{i-5:i-1}) \bigg)
\end{align}
where $\lambda = 1.0$ balances acoustic and language model evidence and $\alpha = 0.9$ is the length normalization exponent. The length normalization $n^\alpha$ with $\alpha < 1$ mildly penalizes longer sequences, counteracting the tendency of log-probability sums to decrease monotonically with sequence length, which would otherwise bias the decoder toward shorter (and therefore less informative) hypotheses.

\begin{figure}[H]
\centering
\includegraphics[width=0.9\linewidth]{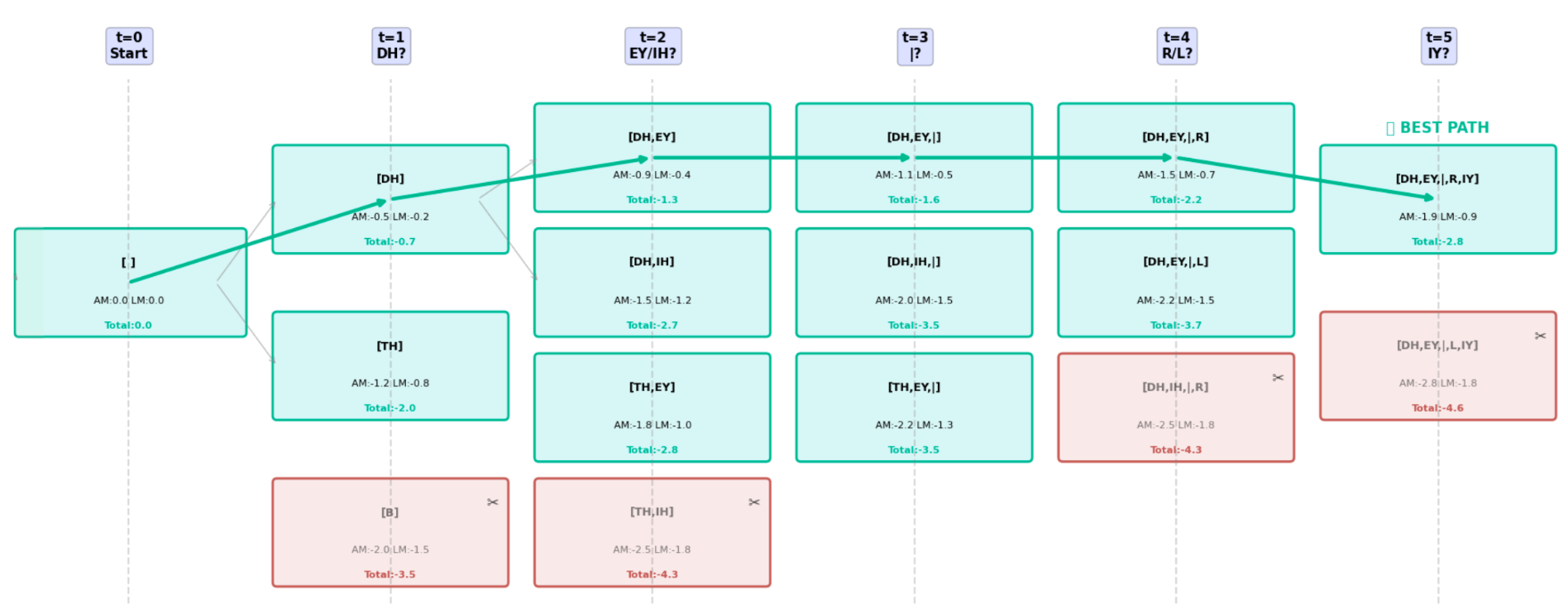}
\caption{Beam search visualization (simplified to width 3). Hypotheses expand, score, and prune at each step. Red: pruned; green: retained best path.}
\label{fig:beam_search}
\end{figure}

Beam search with width $K = 128$ maintains the top-$K$ partial hypotheses at each of the $T/8$ subsampled time steps. At each step, every hypothesis is expanded by all 42 possible next tokens (41 phonemes + blank), yielding up to $42K$ candidates. Each candidate is scored through the WFST using the combined acoustic-linguistic objective, and only the top-$K$ scoring paths are retained. This pruning strategy reduces the $O(42^{T/8})$ exhaustive search to $O(42 \cdot K \cdot T/8)$ operations per sequence---for a typical 3-second trial with $T/8 = 19$ frames and $K = 128$, this amounts to approximately 102,000 score evaluations rather than $42^{19} \approx 10^{30}$. The optimal sequence is selected at the final time step: $y^* = \arg\max_y \text{Score}(y)$.


\begin{table*}[!t]
\centering
\caption{Decoding Approach: Standard Methods vs. Our BCI Application}
\label{tab:decoding_novelty}
\begin{tabular}{@{}llll@{}}
\toprule
\textbf{Standard Method} & \textbf{Conventional Usage} & \textbf{Our Application} & \textbf{Origin} \\
\midrule
Beam search algorithm & Greedy CTC decoding in ASR & Applied to BCI/neural decoding without acoustic dictionary & Freitag (2017) \\
N-gram language model & Word-level LM or none & Phoneme-level 6-gram with Kneser-Ney + confusion matrix & Freitag / Chen~\cite{chen1999empirical} \\
CTC decoding output & CTC $\to$ greedy argmax & Combined with phoneme LM rescoring via recursive backoff & Graves~\cite{graves2006connectionist} \\
Length normalization & Fixed penalty or none & Bayesian optimization via Optuna (150 trials) & Akiba~\cite{akiba2019optuna} \\
WFST graph structure & Word-segmented decoding & Continuous speech without word boundaries & Mohri~\cite{mohri2002weighted} \\
\bottomrule
\end{tabular}
\end{table*}

Table~\ref{tab:decoding_novelty} contextualizes our decoding approach relative to prior methods. The novel idea is applying beam search through a WFST graph with phoneme-level (not word-level) language modeling to BCI neural decoding. Essentially, we have adapted techniques from ASR~\cite{mohri2002weighted,heafield2011kenlm} to a domain where no acoustic dictionary exists, requiring the LM and WFST to jointly create a T15-specific ``phoneme dictionary'' learned from data.

\section{Clinical Application and User Interface}

Eye-tracking is the primary alternative input modality for individuals with severe motor impairment, but gaze-based interfaces suffer from the Midas touch problem first identified by Jacob~\cite{jacob1990eye}: because the eye serves simultaneously as perceptual organ and pointing device, every fixation during natural visual scanning is a potential unintended click. This manifests as two distinct failure modes: (a) low spatial accuracy from physiological gaze jitter (typically 0.5--1.0\textdegree{} visual angle, corresponding to 15--30 pixels on a typical display at 60~cm viewing distance) causes selection of adjacent targets, and (b) saccadic eye movements through intermediate targets during gaze travel trigger unintended activations of passed-over elements. Current solutions rely on dwell-time thresholds~\cite{majaranta2006twenty}---typically 500--1000~ms of sustained fixation to confirm selection---which cap communication speed at approximately 5--10 words per minute and introduce substantial user fatigue from the sustained concentration required.

The fundamental issue is that conventional gaze interaction conflates two logically distinct operations: \textit{pointing} (directing attention to a target) and \textit{selection} (confirming intent to act on that target). In mouse-based interaction, these are naturally separated: the mouse moves the cursor (pointing) while the button click confirms (selection). Fitts' Law~\cite{fitts1954information} predicts that pointing time $T = a + b \cdot \log_2(D/W + 1)$, where $D$ is distance and $W$ is target width, applies to gaze pointing with coefficients that reflect the eye's saccadic dynamics rather than hand motor control. However, for selection, no analogous physical separation exists with eye-only input---the dwell-time mechanism is a temporal proxy that fundamentally constrains throughput.

We propose ``iPhoneme: Slide to Unlock''---a chorded input paradigm that restores the pointing-selection separation by using silent speech phoneme detection via iEEG as the selection channel while gaze serves exclusively as the pointing channel. The key insight is that BCI-enabled patients who have iEEG implants for speech decoding possess a secondary input modality---intentional phoneme production---that is completely independent of eye movement and can therefore serve as an orthogonal confirmation signal. This is analogous to keyboard modifier keys (Ctrl, Shift) that transform the meaning of a simultaneous keystroke; our approach transforms a passive gaze fixation into an active selection by requiring concurrent phoneme production.

This chorded input provides three properties that dwell-time cannot achieve. First, \textit{intentionality}: accidental fixations during natural visual scanning do not trigger selection because no phoneme is being simultaneously produced; the user must deliberately engage both channels. Second, \textit{safety-critical control}: actions with serious consequences (deleting text, sending messages, making medical decisions) can require specific high-confidence trigger phonemes, providing a safety layer analogous to ``confirm'' dialogs but without the cognitive overhead of navigating to a button. Third, \textit{speed}: phoneme detection latency from our ConformerXL model is approximately 100--180~ms, which is 3--10$\times$ faster than typical dwell-time thresholds of 500--1000~ms, directly increasing communication bandwidth.

\begin{figure}[!t]
\centering
\includegraphics[width=0.95\linewidth]{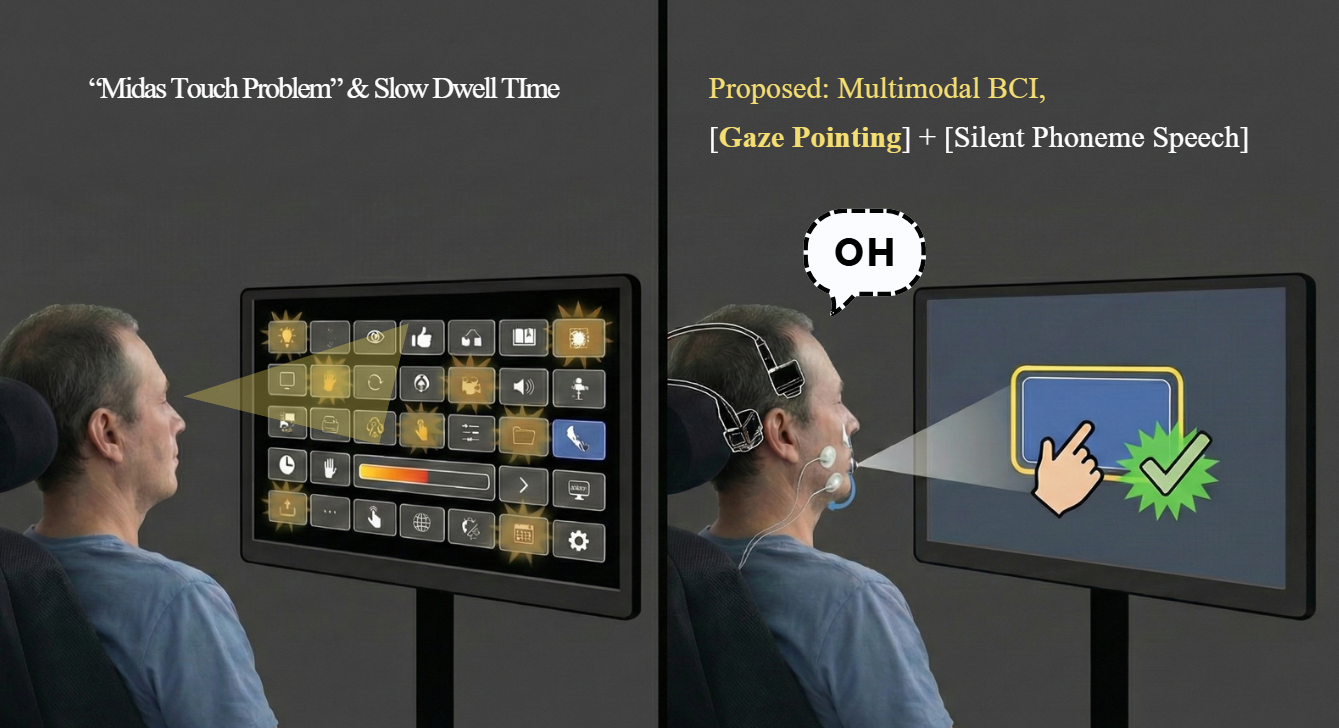}
\caption{iPhoneme: Slide to Unlock. Gaze swipe + silent speech phoneme trigger replaces dwell-time, providing chorded intentional selection.}
\label{fig:iphoneme}
\end{figure}

We define two interaction modalities that extend the chorded paradigm beyond simple click-replacement. In gaze-phoneme swipe, the user sustains a trigger phoneme (e.g., ``Oh'') while sweeping their gaze in a direction; the phoneme onset as detected by the iEEG decoder marks gesture start, the gaze trajectory during the sustained phoneme window determines the swipe vector (direction and distance), and phoneme offset marks gesture end. The temporal binding between phoneme production and gaze movement creates a gesture that is far more expressive than a dwell-time click. This can encode direction, distance, and speed, enabling page-turning, screen transitions, scrolling, and spatial navigation commands. 

In gaze-phoneme text drag, a different trigger phoneme (e.g., ``Ah'') initiates a text selection mode: phoneme onset anchors the selection start at the current gaze position, gaze movement extends the selection highlight across text, and phoneme offset finalizes the selection and triggers a copy action. This enables text selection, copy, and paste operations---interactions that are extremely difficult with dwell-time-based gaze interfaces because they require maintaining state across a continuous gesture rather than discrete point selections.

\begin{figure}[!t]
\centering
\includegraphics[width=0.95\linewidth]{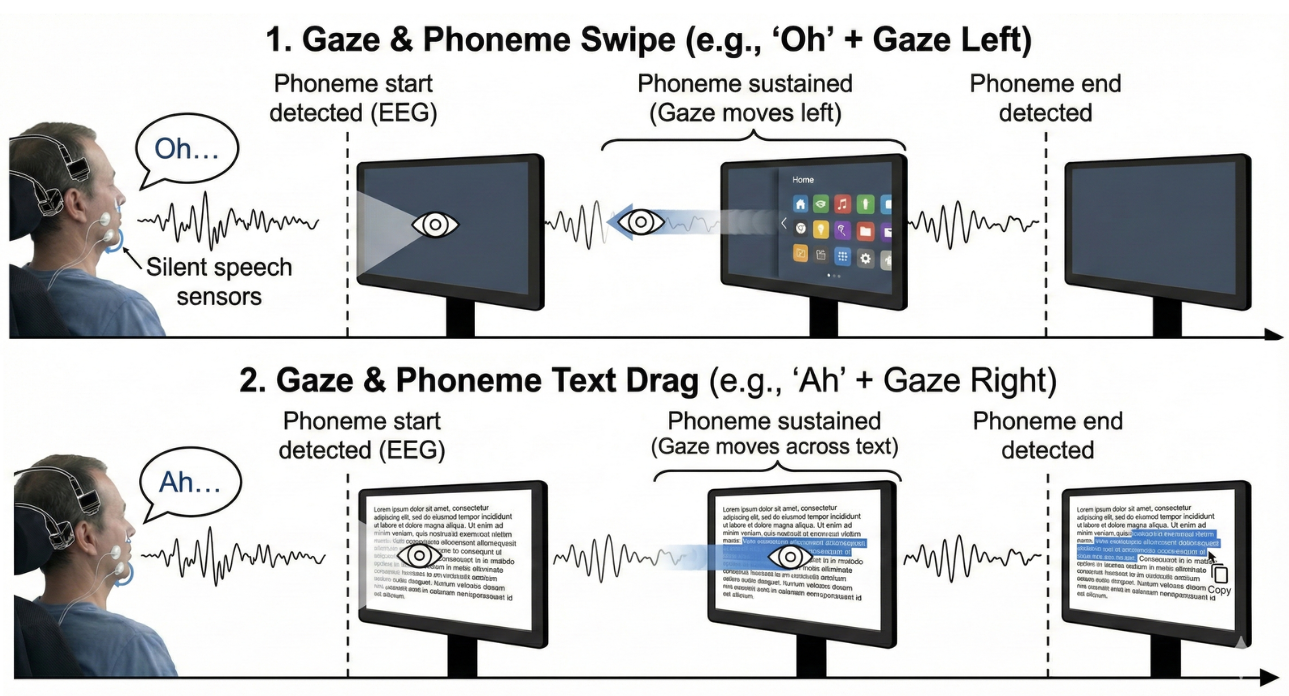}
\caption{Gaze-phoneme interaction modalities. Top: swipe navigation. Bottom: text drag-and-copy.}
\label{fig:gaze_phoneme}
\end{figure}
Choosing the appropriate trigger phonemes is a non-trivial optimization problem balancing three competing objectives: detection accuracy (to minimize false rejections that frustrate the user), natural speech rarity (to minimize false activations during conversation), and articulatory distinctiveness (to ensure the phoneme can be reliably produced even by patients with partial motor impairment). We formalize this as a composite score:
\begin{equation}
    \text{TriggerScore}(p) = \frac{\text{Precision}(p) \times \text{Recall}(p)}{\text{Frequency}(p) + \epsilon}
\end{equation}
where $\epsilon = 0.001$ prevents division by zero for extremely rare phonemes, Precision and Recall are computed from our ConformerXL's per-phoneme confusion matrix on the validation set, and Frequency is the phoneme's occurrence rate in natural English speech. The numerator is the F1-like product of detection reliability, while the denominator penalizes phonemes that occur frequently in conversation (which would produce frequent false activations when the patient is using the speech decoding mode rather than the UI control mode).
 
\begin{figure}[!t]
\centering
\includegraphics[width=0.95\linewidth]{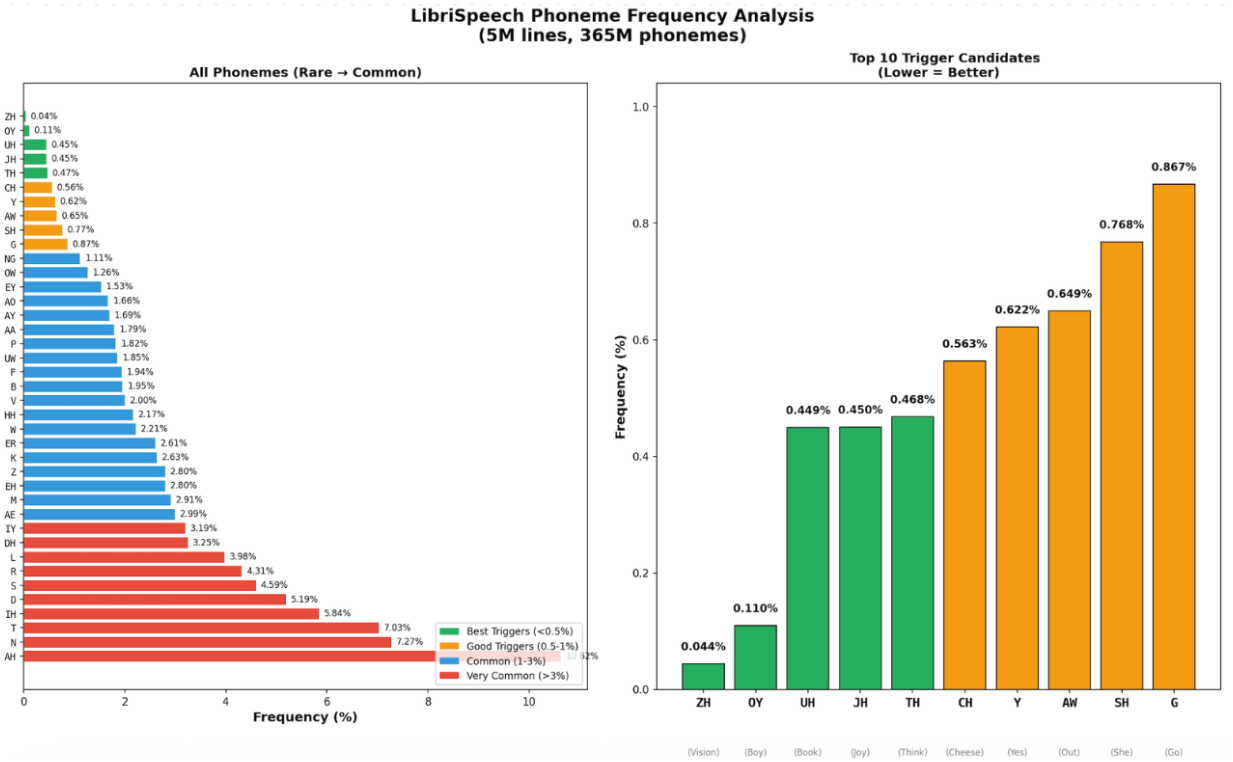}
\caption{Phoneme frequency analysis from LibriSpeech (365M phonemes). Left: frequency ranking. Right: top-10 trigger candidates by rarity.}
\label{fig:frequency}
\end{figure}
 
\begin{figure}[!t]
\centering
\includegraphics[width=0.95\linewidth]{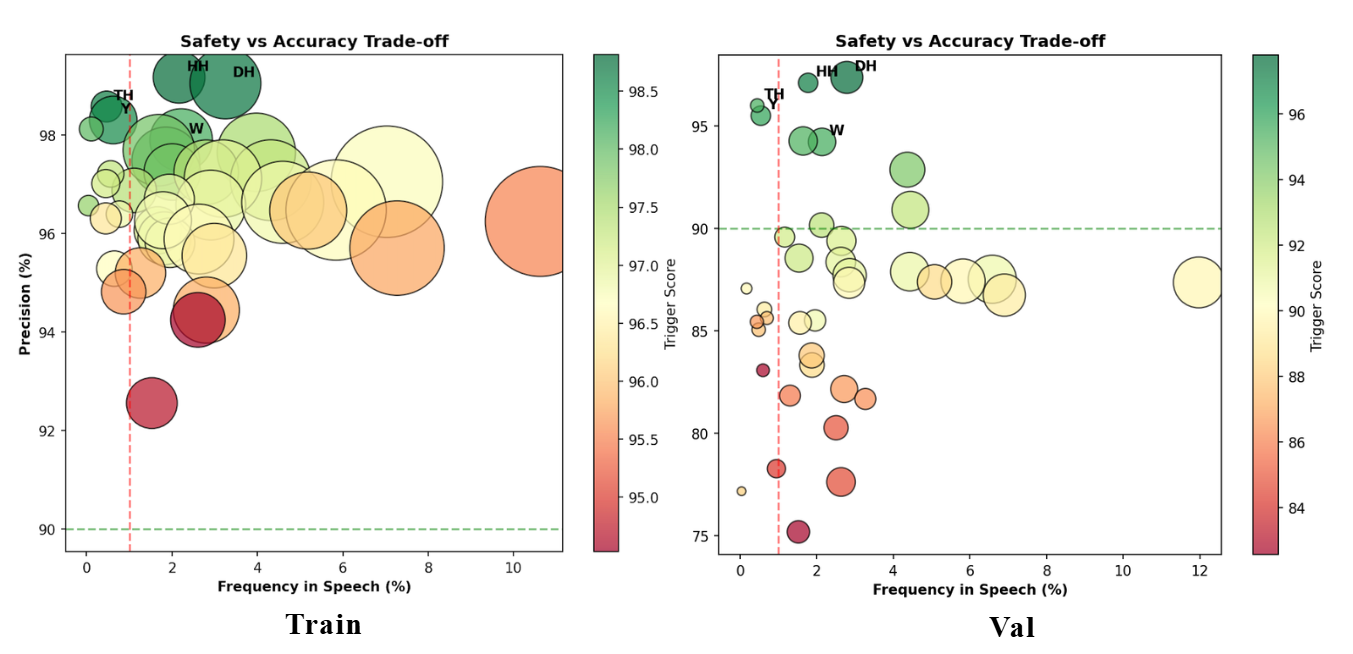}
\caption{Safety vs. accuracy trade-off. Ideal triggers (upper-left) combine high precision with low frequency. DH, HH, TH, Y, W consistently occupy the optimal region.}
\label{fig:trigger_tradeoff}
\end{figure}

Analysis of phoneme frequencies from LibriSpeech (5M utterances, 365M total phonemes) identifies ZH (0.04\%), OY (0.11\%), UH (0.45\%), JH (0.45\%), and TH (0.47\%) as the rarest phonemes. However, rarity alone is insufficient---a rare phoneme that is frequently confused with common phonemes would still produce false activations. Combining frequency with our per-phoneme precision and recall measurements, the top-5 trigger candidates are DH (score: 97.2), HH (95.8), Y (94.1), W (93.5), and TH (92.9), all of which occupy the desirable upper-left quadrant of the safety-accuracy trade-off space (Fig.~\ref{fig:trigger_tradeoff}): high precision ($>$94\% on validation) and low natural frequency ($<$3.5\% of speech phonemes).
 
\section{Results}

Table~\ref{tab:main_results} summarizes decoding performance across all pipeline stages. The WFST beam search yields +2.76\% absolute improvement over greedy decoding (89.38\%$\to$92.14\%), with error breakdown of 5.2\% substitutions, 1.8\% deletions, and 0.86\% insertions. The dominance of substitution errors (66\% of total errors) over deletions (23\%) and insertions (11\%) indicates that the acoustic model generally identifies the correct number of phonemes but sometimes selects the wrong class---a pattern consistent with articulatory confusion between similar phonemes rather than fundamental timing or segmentation failures. At the word level, we achieve 73.39\% accuracy (26.61\% WER), where the gap between phoneme and word accuracy reflects the compounding effect: a word of average length $\bar{L} = 4.5$ phonemes has expected correctness $(1 - 0.0786)^{4.5} \approx 0.693$ under independent errors, and our observed 73.39\% slightly exceeds this, suggesting the WFST corrects correlated within-word errors. Qualitatively, early-session (2023) trials such as ``Bring it closer'' and ``What do they like?'' are decoded with near-perfect accuracy, while late-session (2025) trials containing terms like ``barometric'' or brand names produce the majority of word-level errors due to phoneme sequences unseen in the LM training data.

\begin{table}[!t]
\centering
\caption{Decoding Performance Across Pipeline Stages}
\label{tab:main_results}
\begin{tabular}{@{}lccc@{}}
\toprule
\textbf{Method} & \textbf{PER (\%)} & \textbf{Acc. (\%)} & \textbf{Latency} \\
\midrule
Greedy (argmax) & 10.62 & 89.38 & 45 ms \\
+ 6-gram LM (0.8) & 9.98 & 90.02 & 65 ms \\
+ WFST beam ($K$=128) & \textbf{7.86} & \textbf{92.14} & 180 ms \\
\bottomrule
\end{tabular}
\end{table}

Table~\ref{tab:beam_sweep} shows the top-4 decoder configurations from 150 Optuna trials. All top configurations converge on LM weight 1.0 and length penalty 0.9--1.0, suggesting that equal weighting of acoustic and language model evidence is optimal for this task. Beam sizes $\geq$64 consistently achieve $>$92\%, with diminishing returns beyond 128.

\begin{table}[!t]
\centering
\caption{WFST Beam Search Top-4 (150 Optuna Trials)}
\label{tab:beam_sweep}
\begin{tabular}{@{}cccc@{}}
\toprule
\textbf{Beam} & \textbf{LM Wt.} & \textbf{Len. Pen.} & \textbf{Acc. (\%)} \\
\midrule
128 & 1.0 & 0.9 & \textbf{92.14} \\
96  & 1.0 & 0.9 & 92.11 \\
128 & 1.0 & 1.0 & 92.08 \\
64  & 1.0 & 0.9 & 92.06 \\
\bottomrule
\end{tabular}
\end{table}

Table~\ref{tab:comparison} situates our results within the broader literature. Our system achieves approximately 3\% absolute improvement over Card et al.~\cite{card2024accurate} (89\%), the prior state-of-the-art on the same T15 dataset.

\begin{table}[!t]
\centering
\caption{Comparison with Prior Work}
\label{tab:comparison}
\begin{tabular}{@{}lcc@{}}
\toprule
\textbf{Study} & \textbf{Year} & \textbf{Acc. (\%)} \\
\midrule
Wairagkar et al.~\cite{wairagkar2024} & 2024 & 56 \\
Willett et al.~\cite{willett2023speech} & 2024 & 82 \\
Card et al.~\cite{card2024accurate} & 2025 & 89 \\
\textbf{This Work} & \textbf{2025} & \textbf{92.14} \\
\bottomrule
\end{tabular}
\end{table}

\begin{figure}[!t]
\centering
\includegraphics[width=0.95\linewidth]{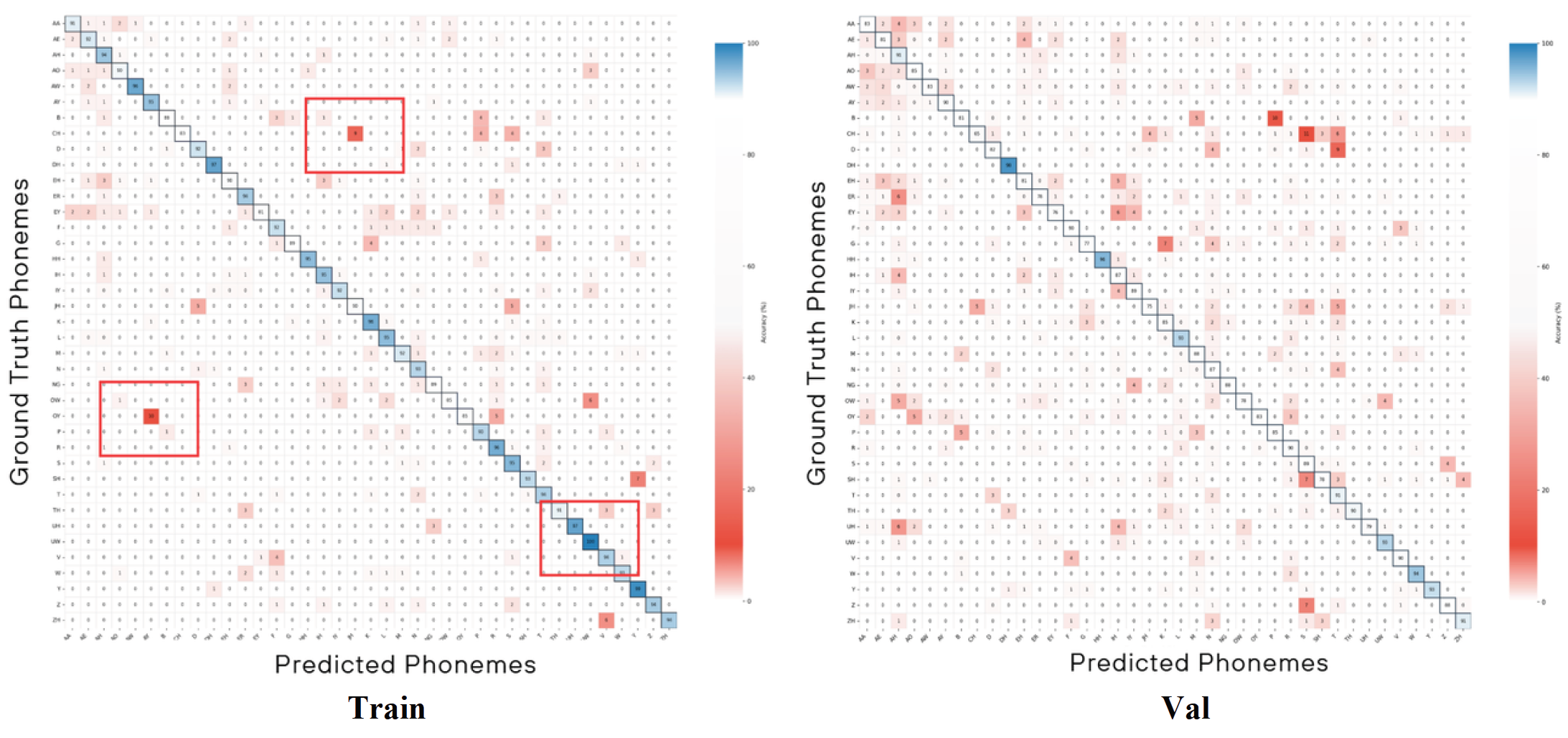}
\caption{Per-phoneme confusion matrices (train left, validation right). Red boxes highlight confusion clusters among articulatorily similar phonemes.}
\label{fig:confusion_matrix}
\end{figure}

\begin{figure}[!t]
\centering
\includegraphics[width=0.95\linewidth]{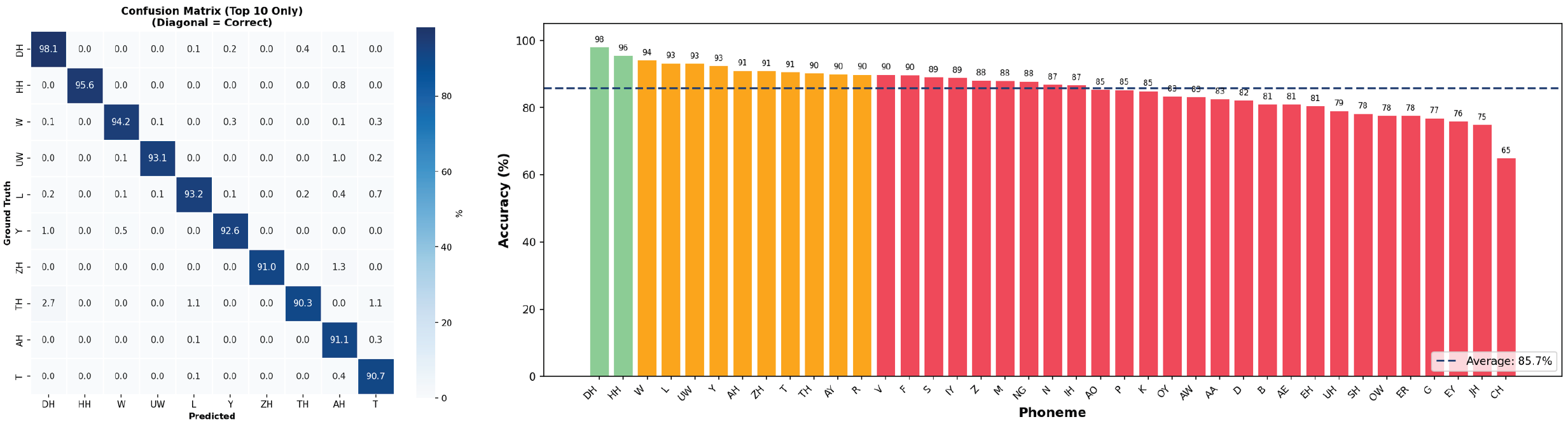}
\caption{Left: per-phoneme accuracy ranking (avg. 85.7\%). Right: trigger candidate scores with top-5 (DH, HH, Y, W, TH) highlighted.}
\label{fig:per_phoneme}
\end{figure}

Fig.~\ref{fig:confusion_matrix} presents the full 41$\times$41 phoneme confusion matrices for both training and validation sets. The strong diagonal indicates high per-class accuracy, while off-diagonal clusters reveal systematic confusions among articulatorily similar phonemes. The top-10 most accurate phonemes on validation are DH (98.1\%), HH (95.6\%), W (94.2\%), L (93.2\%), UW (93.1\%), Y (92.6\%), AH (91.1\%), ZH (91.0\%), TH (90.3\%), and T (90.7\%). The most confused phonemes---CH (65\%), JH (75\%), EY (76\%), and G (77\%)---share articulatory features with multiple other classes, consistent with the known overlap in motor cortex representations for similar articulatory gestures~\cite{bouchard2013functional}. Fig.~\ref{fig:per_phoneme} shows the per-phoneme accuracy ranking and trigger score analysis.

\section{Discussion}

The 92.14\% phoneme accuracy and 73.39\% word accuracy represent the highest reported performance on the T15 dataset. The gap between phoneme and word accuracy---18.75 percentage points---reflects the compounding nature of phoneme errors: a word of average length 4.5 phonemes requires all constituent phonemes to be correct for the word to be scored as accurate, so a per-phoneme error rate of 7.86\% translates to an expected word correctness of approximately $(1 - 0.0786)^{4.5} \approx 0.693$, or 69.3\%. Our observed 73.39\% exceeds this prediction, suggesting that the WFST beam search corrects correlated phoneme errors within words more effectively than the independent-error assumption would predict---likely because the 6-gram LM captures within-word phonotactic patterns that constrain multiple phonemes simultaneously.

The architectural decisions---8$\times$ GELU subsampling for CTC stability, the multi-scale dilated convolution + BiGRU temporal prenet for neural jitter correction, and Pre-RMSNorm for training stability on non-stationary biological signals---each address specific failure modes of standard Conformer architectures when applied to iEEG rather than audio. The WFST beam search with properly implemented recursive backoff LM adds a further +2.76\% over greedy decoding by enforcing English phonotactic constraints during inference. Notably, this improvement was achieved only after resolving a subtle implementation issue in the backoff mechanism: our initial LM integration assigned uniform probability $1/42$ to unseen n-grams rather than recursively backing off to shorter contexts. This originally eliminated the LM's discriminative power and caused the decoder to perform worse than greedy.

The per-phoneme analysis reveals that confusion patterns are strongly correlated with articulatory similarity. The most confused phonemes---CH (65\%), JH (75\%), EY (76\%), G (77\%)---share manner or place of articulation with multiple other classes. For instance, CH and JH differ only in voicing, which produces minimal differences in motor cortex activity since voicing is primarily controlled by laryngeal muscles rather than the cortical regions sampled by our electrodes~\cite{bouchard2013functional}. This suggests that future accuracy improvements may require either additional electrode coverage of laryngeal motor cortex or direct modeling of articulatory feature sharing in the decoder architecture.

However, several limitations temper these results. The BiGRU's bidirectional processing introduces 100--200~ms latency overhead that would compound in real-time streaming applications, as the backward pass requires the full input sequence to be available before computation can begin. For deployment, the input would need to be segmented into overlapping windows of 500--1000~ms with BiGRU applied per-window, sacrificing some cross-window context at boundaries. 

Furthermore, validation accuracy drops from near-perfect on 2023 sessions to approximately 80.8\% on 2025 sessions, a degradation driven by both vocabulary difficulty and the dynamic nature of neural signals. As Hodgkin and Huxley established in their foundational model of neuronal dynamics, membrane potential behavior depends on time-varying Na$^+$/K$^+$ ion channel conductances that change with aging, disease progression, and circadian state, meaning the mapping from cortical activity to phoneme intent is inherently non-stationary. The 7-month recording gap between the 2024 and 2025 data exacerbates this, as the model has no training signal to track the gradual drift during the gap period. Future work should explore continual learning approaches, session-adaptive normalization, or temporally-weighted training that assigns higher importance to more recent sessions.

The approximately 10~GB training data is also likely a bottleneck. Modern ASR systems train on thousands of hours of speech data (LibriSpeech alone contains 960 hours), while our 8,071 trials at approximately 3 seconds each total only about 6.7 hours of neural recordings. This constrains model capacity---larger models with more parameters tend to overfit on this data volume---and limits the effectiveness of data augmentation strategies. 

Finally, all results are from a single participant (T15). Cross-participant generalization would require either substantial transfer learning techniques or participant-specific calibration protocols, as the spatial mapping from electrode positions to cortical function varies considerably across individuals.

\section{Future Directions}

For future directions, we identify three primary paths addressing the above limitations. The first is developing a custom phoneme-to-sentence language model to replace the Claude Sonnet 4.5 dependency, potentially using a small transformer trained on phoneme-word aligned pairs from CMU Dictionary and LibriSpeech, which would reduce inference cost and enable fully offline operation. 

The second is exploring novel decoding architectures. We designed a preliminary Stochastic Trajectory Graph (STG) decoder: learned and dynamic edge potentials (vs. fixed in HMM/Viterbi or compiled in WFST), $K$-iteration message-passing inference (vs. single forward pass), belief-based path communication between hypotheses (vs. independent paths in beam search), full-sequence global context (vs. limited beam context), input-dependent adaptivity (vs. fixed structure), and full posterior uncertainty estimation (vs. point estimates or top-$K$ lists). However, quick testing showed the conventional WFST + beam structure outperformed STG on this dataset, possibly because the STG's increased expressivity introduces optimization difficulties that outweigh its representational advantages at the current data scale. We report the WFST results and leave STG development as future work.

The third is finding alternative inputs modalities beyond eye-gaze entirely. 
Augmental's MouthPad \cite{augmental2023mouthpad} demonstrates intraoral interfaces are clinically viable, mapping tongue-to-palate contact to 2D 
cursor control. However, beyond that, a more expressive extension could be treating the full 3D volumetric 
space of the oral cavity as a continuous pointing space, where the tongue's position along the left-right, anterior-posterior, and superior-inferior axes maps directly to cursor coordinates. This may reduce even more ocular fatigue while providing greater degrees of freedom than any 2D surface interface. Combined with iPhoneme's silent  speech confirmation channel, such a system would yield a fully hands-free, gaze-free 
interaction. Nevertheless, these of miniaturized 3D tongue-position sensors for chronic intraoral use remains an open engineering challenge.


\begin{table}[H]
\centering
\footnotesize
\caption{Ethical Impact Analysis}
\label{tab:ethics}
\begin{tabular}{@{}p{2cm}p{2.7cm}p{2.7cm}@{}}
\toprule
\textbf{Component} & \textbf{Challenge} & \textbf{Mitigation} \\
\midrule
Accuracy (+3\% PER) & Phoneme hallucinations may misrepresent intent in medical/legal contexts & Confidence gating; top-$k$ candidates shown for correction \\
\addlinespace
ConformerXL + WFST & Large models fail out-of-distribution; latency harms daily use & WFST+LM gives stable behavior; CPU deployment removes GPU dependency \\
\addlinespace
Gaze + phoneme UI & Decoding errors erode trust in cognitively intact but physically dependent users & Confidence visualization; chorded input prevents accidental actions \\
\addlinespace
Eye-tracking + iEEG & One-size UX excludes users with varying neural profiles or disease progression & Adaptive LM weighting; personalized trigger calibration per session \\
\bottomrule
\end{tabular}
\end{table}

\section{Ethical Considerations}

Table~\ref{tab:ethics} maps each system component to its associated ethical challenges and proposed mitigations. The core tension is between maximizing accuracy (which risks hallucinated phonemes that misrepresent user intent) and maintaining user trust and safety (which requires transparent confidence indicators and correctable outputs). Our multimodal interface design addresses this by providing top-$k$ candidate corrections and requiring deliberate chorded input for important actions.

\balance
\section{Conclusion}

We presented a integrated brain-to-text communication system for ALS patients achieving 92.14\% phoneme accuracy (7.86\% PER) and 73.39\% word accuracy (26.61\% WER) on the T15 intracranial EEG dataset---approximately 3\% above prior state-of-the-art---through a ConformerXL acoustic model with novel temporal BiGRU prenet, 8$\times$ GELU subsampling, and Pre-RMSNorm, combined with WFST beam search and a 6-gram phoneme LM tuned over 150 Optuna trials. As part of the system, the ``iPhoneme'' interface introduces chorded gaze-plus-phoneme input that formally separates the pointing and selection channels in gaze interaction, solving the Midas touch problem without the speed penalty of dwell-time thresholds.

The system highlights three key design principles that may generalize beyond this specific application. First, that Conformer-based architectures designed for audio ASR can be effectively adapted for neural signal decoding through targeted modifications to the prenet, subsampling, and normalization layers, rather than requiring entirely novel architectures. Second, the combination of WFST beam search with phoneme-level n-gram language models---a well-established technique in ASR---provides gains when applied to BCI decoding, where the absence of an acoustic dictionary requires the LM to become both phonotactic constraint and implicit lexicon. Third, that the Midas touch problem in gaze interaction can be addressed through multimodal chorded input that leverages the BCI implant's existing phoneme detection capability as a secondary confirmation channel. This effectively creates new interaction modalities (swipe, text drag) that are fundamentally inaccessible to single-modality gaze systems.

The system deploys on CPU at 180~ms latency. Future work targets the 95\% clinical accuracy threshold through ensemble methods that combine multiple ConformerXL models with diverse hyperparameters, neural language models that capture longer-range context than 6-gram statistics, temporal domain adaptation through session-weighted training and continual learning, and clinical validation with ALS patients in realistic scenarios, toward enabling practical, real-time communication in everyday settings.

\bibliographystyle{IEEEtran}

\end{document}